\documentclass{aa}
\usepackage[varg]{txfonts}

\usepackage[breaklinks, colorlinks, citecolor=blue, urlcolor = blue]{hyperref}
      
\usepackage{csquotes}
\usepackage{units}
\usepackage{mathtools}
\usepackage{float}
\usepackage[caption = false]{subfig}
\usepackage{graphicx}
\usepackage{amsmath}

\bibpunct{(}{)}{;}{a}{}{,}

\begin{document}

\title{Variability of the soft X-ray excess in IRAS\,13224--3809}
\author{E. S. Kammoun\inst{\ref{inst1},\ref{inst2}\thanks{Former graduate student at the Department of Physics \& Astronomy, Notre Dame University-Louaize, Zouk Mosbeh, Lebanon}}
\and I. E. Papadakis\inst{\ref{inst3},\ref{inst4}}
\and B. M. Sabra\inst{\ref{inst2}}
}

\institute{Astrophysics Sector, SISSA, Via Bonomea 265, I--34136, Trieste, Italy; \email{\href{mailto:ekammoun@sissa.it}{ekammoun@sissa.it}}\label{inst1}
\and
Department of Physics \& Astronomy, Notre Dame University-Louaize, P.O. Box 72 Zouk Mikael, Zouk Mosbeh, Lebanon\label{inst2}
\and
Department of Physics and Institute of Theoretical and Computational Physics, University of Crete, 71003 Heraklion, Greece\label{inst3}
\and
IESL, Foundation for Research and Technology, 71110 Heraklion, Greece\label{inst4}
}

\date{Received 23 June 2015 / Accepted 10 August 2015}
\abstract
{ We study the soft excess variability of the narrow line Seyfert 1 galaxy IRAS\,13224--3809. We considered all five archival {\it XMM-Newton} observations, and we applied the `flux--flux plot' (FFP) method. We found that the flux--flux plots were highly affected by the choice of the light curves' time bin size,  most probably because of the fast and large amplitude variations, and the intrinsic non-linear flux--flux relations in this source. Therefore, we recommend that the smallest bin--size should be used in such cases. Hence, we constructed FFPs in 11 energy bands below 1.7 keV, and we considered the 1.7--3 keV band, as being representative of the primary emission. The FFPs are reasonably well fitted by a `power-law plus a constant' model. We detected significant positive constants in three out of five observations. The best-fit slopes are flatter than unity  at energies below $\sim 0.9$ keV, where the soft excess is strongest. This suggests the presence of intrinsic spectral variability. A power--law-like  primary component, which is variable in flux and spectral slope (as $\Gamma\propto N_{\rm PL}^{0.1}$) and a soft-excess component, which varies with the primary continuum (as $F_{\rm excess}\propto F_{\rm primary}^{0.46}$),  can broadly explain the FFPs. In fact, this can create positive `constants', even when a stable spectral component does not exist. Nevertheless, the possibility of a stable, soft--band constant component cannot be ruled out, but its contribution to the observed 0.2--1 keV band flux should be less than $\sim 15$\%. The model constants in the FFPs were consistent with zero in one observation, and negative at energies below 1\,keV in another.  It is hard to explain these results in the context of any spectral variability scenario, but they may signify the presence of a variable, warm absorber in the source.}


\maketitle
\section{Introduction}
\label{sec:intro}

It is generally accepted that AGN are powered by accretion of matter onto a central supermassive black hole (SMBH) of mass $M_{\rm BH}\sim 10^{6-9}{\rm M_{\odot}}$. The matter is thought to accrete in a disc that is geometrically thin and optically thick  \citep{Shak73}. If the released gravitational energy is dissipated locally on the disc, then we expect a multi-temperature, blackbody emission component to dominate the spectral energy distribution of these objects, with a maximum temperature of  $\sim 10^5{\rm K}$ that peaks in the ultraviolet. Indeed, observations of AGN suggest the presence of a peak in the optical-UV continuum known as the `big blue bump' \citep{Mal82,Lao97,Zhe97,Sha05}.

AGN are strong X-ray emitters. These X-rays are believed to be triggered by Compton up-scattering of the disc photons off hot electrons surrounding the disc, in a hot ($\sim 10^9\,{\rm K}$)  medium, usually referred to as the X-ray corona. The X-ray photons are emitted isotropically, and although the geometry of disc-corona  is currently unknown in AGN, they may reflect off the accretion disc and/or the dusty torus (0.1--10 pc from the SMBH) , thereby producing an additional `reflection' component in the X-ray spectrum \citep[e.g.][]{Geo91}. 

AGN X-ray spectra at energies above $\sim 2$\,keV have a power-law-like shape. The extrapolation of the best-fit  power-law models  to energies lower than 2 keV reveals in many AGNs a spectral component in excess of the extrapolated hard-band continuum. This excess emission is know as the soft X-ray excess. This component was discovered over 30 years ago \citep{Sin85,Arn85}, and its source has been debated ever since. 

Originally, it was  suggested that soft X-ray excess represents the high-frequency tail of the disc emission in AGN. In fact, the soft excess in many AGN could be fitted by a blackbody model with a best-fit temperature in the range $0.1-0.2$\, keV \citep{Wal93,Cze03}, however, this temperature is significantly higher than the maximum temperatures expected in AGN accretion discs. It has also been proposed that the soft excess could arise from Compton up-scattering of disc photons, in a `warm' medium of an electron population with a temperature much lower and an optical depth much higher than those of the X-ray corona that are responsible for the emission at energies above 2\,keV  \citep[e.g.][]{Cze87,Mag98,Jani01}. However, \citet{Gie04} show that the temperature associated with this Comptonisation region is constant over a wide range of AGN luminosity and black hole mass, a result that requires a fine-tuning between the optical depth and the heating/cooling ratio of this region. They later argue \citep{Gie06} that a ``smeared'' absorption model can provide an explanation for this component, but subsequent studies show that line-driven AGN accretion disc winds cannot reproduce the soft excess \citep{Sch07, Sch08}. 

Recently, \cite{Don12} have revived the original idea of the soft excess arising from the disc itself. They propose a model where the gravitational energy released by accretion at small radii is split between powering optically thick, Comptonized disc emission, which is responsible for the soft excess, and an optically thin corona above the disc, which is responsible for the high-energy X-ray emission. Finally, X-ray reflection of the inner disc by the X-ray source can also account for the soft excess. If the accretion disc is mildly ionized and the X-rays illuminate its innermost region, excess emission below $\sim 2$\,keV could appear because of line and bremsstrahlung emission from the hot disc layers. \cite{Cru06} explain the soft excess in many AGN using a relativistically-blurred version of the disc reflection model of \cite{Ros05}, but recent studies have shown that this model may not be sufficient in some cases \citep[see][]{Loh12}. 

In this work, we apply the flux--flux plot (FFP) method using archival {\it XMM-Newton} data of the narrow line Seyfert 1 (NLS1) galaxy \object{IRAS\,13224--3809}. Our aim is to study the soft X-ray variability in this source, and constrain the origin of its soft X-ray excess in a model-independent fashion. The FFP method has recently been  applied by \cite{Nod11, Nod13} to a handful of X-ray bright quasars with the same goals. They detect a soft component in the 0.5--3\,keV band of a few AGN, which is less variable than the primary source, and they interpreted this in terms of thermal Comptonization of the disc emission. This method was first developed by \cite{Chu01} to identify variable and stable components in the X-ray spectrum of the black hole binary Cygnus X-1. It was later applied by \cite{Tay03} to study the X-ray spectral variability of X-ray bright Seyferts and to many other cases since. 

The FFP method is straightforward in its implementation and is relatively efficient in detecting spectral components that are less variable than the X-ray primary emission. If present, these components result in positive constants in the FFP plots, which, in general, show a strong positive correlation between the flux in various energy bands (for AGN). The FFP variant of the method is particularly effective when studying the fast spectral variability in AGN on timescales when the flux in various energy bands cannot be accurately determined. 

IRAS $13224-3809$ ($z=0.066$) was first detected in X-rays during the {\it ROSAT} all-sky survey in 1992 as a high-luminous X-ray source with an X-ray luminosity $L_{\rm X}=3\times10^{44}\,{\rm erg/s}$ \citep{Bol93}. IRAS $13224-3809$ is highly variable in X-rays. Its spectrum shows a soft X-ray excess below $\sim 1.5$\,keV \citep{Bol96, Bol03} and strong relativistic effects \citep{Pon10,Fab13}. \cite{Chi15} recently presented a detailed study of the X-ray spectral variability of the source. They model its X-ray spectrum in the context of relativistic disc reflection models. They found that the reflected emission is much less variable than the X-ray continuum, and as the source flux drops, the spectrum becomes progressively more reflection-dominated. The X-ray reflection interpretation is broadly consistent with the results of \citet{Emma14}, who studied the soft-band time lags in this source. They found that these are consistent with the hypothesis of a  point--like source which is located at a height of $\sim 3$ gravitational radii (${\rm r_g}$) above a $\sim 10^7$ M$_{\odot}$ black hole (BH). Both the spectral and the timing studies mentioned above suggest that most of the soft X-ray emission in this source is due to X-ray illumination of the inner disc. 

In this paper, we use all the archival {\it XMM-Newton} data of IRAS 13224--3809. The fact that it is quite bright and highly variable in X-rays,  as well as the strong relativistic effects seen in this source, make it an interesting target to apply the FFP method and study its soft X-ray variability. In Section \ref{sec:datared} we present the observations and the data reduction, and in Section \ref{sec:fluxflux} we present the results from the model-fitting of the FFPs. Finally, in the last section we provide a brief summary of our main results and discuss their physical implications.

\section{Observations and data reduction}
\label{sec:datared}
Five observations of IRAS 13224$-$3809 by the {\it XMM-Newton} satellite \citep{Jans01} are available in the {\it XMM-Newton} Science Archive (XSA). The observation log is listed in Table\,\ref{table:obs-log}. The first observation  was done in 2002 January 19 (Obs. ID 0110890101, hereafter Obs.\,1), and the last four observations from 2011 July 19 to 29 (Obs. IDs 0673580101, 0673580201, 0673580301 and 0673580401, hereafter, Obs.\,2, Obs.\,3, Obs.\,4, and Obs.\,5, respectively). All instruments were working successfully during the five observations. We have considered only the data provided by the EPIC-pn camera \citep{Stru01}. 

The first two observations were taken in full window (FW) imaging mode and the others in large window (LW) imaging mode. We reduced all data using the {\it XMM-Newton} Science Analysis System ({\tt SAS } v.13.0.0) and the latest calibration files. The data were cleaned for high background flares and were selected using the condition $ \text{PATTERN} \leq 4$. The net exposure for each observation is also listed in Table\,\ref{table:obs-log}. Source spectra and light curves were extracted from a circle of radius 40$\arcsec$. The respective background spectra and light curves were extracted from an off-source circular region located on the same CCD chip, with a radius approximately twice that of the source  to ensure a high signal-to-noise ratio. Pileup was checked and found to be negligible in all observations. Background-subtracted light curves were produced using the {\tt SAS} task {\tt EPICLCCORR} for different energy bands and time bin sizes. The choice of the energy bands and the time binning is discussed in detail in the next section.

Response matrices were produced using the {\tt SAS} tasks {\tt RMFGEN} and {\tt ARFGEN}. Spectral model-fitting was performed with {\tt XSPEC} v.12.8.1 \citep{Arn96}. We report the $1\sigma$ errors on the model parameter and flux estimates, and $3\sigma$ upper/lower limits when relevant.

\begin{table}
\centering
\caption{Log of observations of IRAS $13224-3809$. The last column lists the 1.7-3\,keV band flux, assuming a power-law model, taking  the Galactic absorption into consideration.}
\begin{tabular}{lccc}
\hline \hline
Obs./date       &       Net exp.        &       Pn-mode & 1.7-3\,keV flux\\     
                                     &       (ks)                      &                         &  ($10^{-13}$ erg/s/cm$^{2}$) \\ \hline \\[-0.2cm]
1/2002-01-19    &       61      &       FW                & $1.94^{+0.11}_{-0.16}$ \\[0.2cm]       
2/2011-07-19    &       120     &       FW                & $2.48^{+0.12}_{-0.19}$ \\[0.2cm]       
3/2011-07-21    &       120     &       LW                & $1.76^{+0.11}_{-0.18}$ \\[0.2cm]
4/2011-07-25    &       110     &       LW                & $0.94^{+0.10}_{-0.14}$ \\[0.2cm]       
5/2011-07-29    &       120     &       LW                & $1.95^{+0.10}_{-0.12}$ \\[0.2cm]       \hline \hline
\end{tabular}
\label{table:obs-log}
\end{table}

\section{Flux--Flux analysis}
\label{sec:fluxflux}

\subsection{The `continuum' band}
\label{subsec:continuum}

\begin{figure}[!ht]
\centering
\includegraphics[scale=0.5]{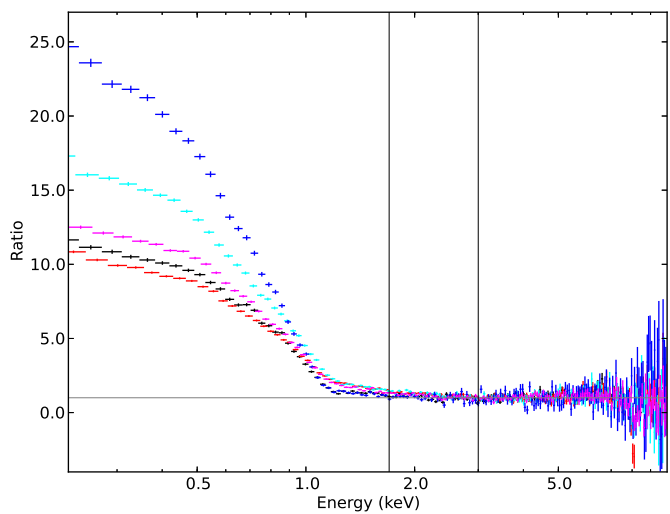}
\caption{Ratio of the observed spectra over a power-law model fitted to the $3-10\,\unit{keV}$ band and extrapolated to lower energies (Obs. 1, 2, 3, 4, and 5 spectra are shown as black, red, cyan, blue, and magenta points, respectively). The two vertical black lines indicate the energy band we chose as representative of the X-ray continuum.}
\label{fig:ratio}
\end{figure}

Our first task was to identify the energy band that is the most representative of the X-ray continuum emission. To this end,  we fitted an absorbed power-law (PL) model to the $3-10\,\unit{keV}$ band spectra for all observations (i.e. {\tt wabs$\ast$powerlaw} in {\tt XSPEC} terminology). We considered only Galactic absorption, which for  the line of sight towards IRAS $13224-3809$ corresponds to an equivalent hydrogen column density of N$_{\rm H}=5.34\times 10^{20}\,{\rm cm^{-2}}$ \citep{Kal05}. Figure\,\ref{fig:ratio} shows the ratio of the observed spectra over the absorbed PL model for the full energy band ($0.2-10.0\,\unit{keV}$) in all observations. Based on the resulting $\chi^2$ values, a PL cannot be accepted as a good fit model in the $3-10$\,keV band. Nevertheless, this model provides a reasonably good base-line model to account for the hard X-ray emission in the source. The extrapolation of the best-fit PL to energies below $\sim 1\,\unit{keV}$ clearly reveals a strong soft excess in all observations. Notably, the strength of this component is also variable. 

Based on the plot shown in Fig.\,\ref{fig:ratio}, we chose the  $1.7-3.0\,\unit{keV}$ band as the most representative of the X-ray `continuum' (indicated by the vertical lines in  Fig.\,\ref{fig:ratio}). The choice of the lower energy limit was dictated by the requirement to be as low as possible (to increase the signal-to-noise in the continuum band)  and at the same time to be the least affected by the soft excess emission. The choice of  high energy limit was dictated by the possibility that the X-ray reflection component may strongly affect  the 3--10\,keV band in this source \citep{Fab13}. In any case, the addition of the 3-10\,keV counts in the continuum band does not increase  the count rate significantly in the band we chose, while it does increase  its error rate considerably (because  the source is relatively faint at energies above $\sim 3$\,keV). The last column in Table\,\ref{table:obs-log} lists the average 1.7-3\,keV band flux of each observation, estimated using the best-fit PL model fits to the 3--10\,keV band data. These values are representative of the continuum flux levels in each observation and show that the X-ray continuum had the highest and lowest flux levels during Obs.\,2 and Obs.\,4, respectively. Conversely, Obs.\,2 (red points in Fig.\,\ref{fig:ratio}) shows the `weakest' soft excess, in terms of ratio, while Obs.\,4 (blue points in the same figure) shows the `strongest' soft excess (i.e. highest ratio).

\begin{figure*}[!ht]
\centering
\subfloat{\includegraphics[scale = 0.32]{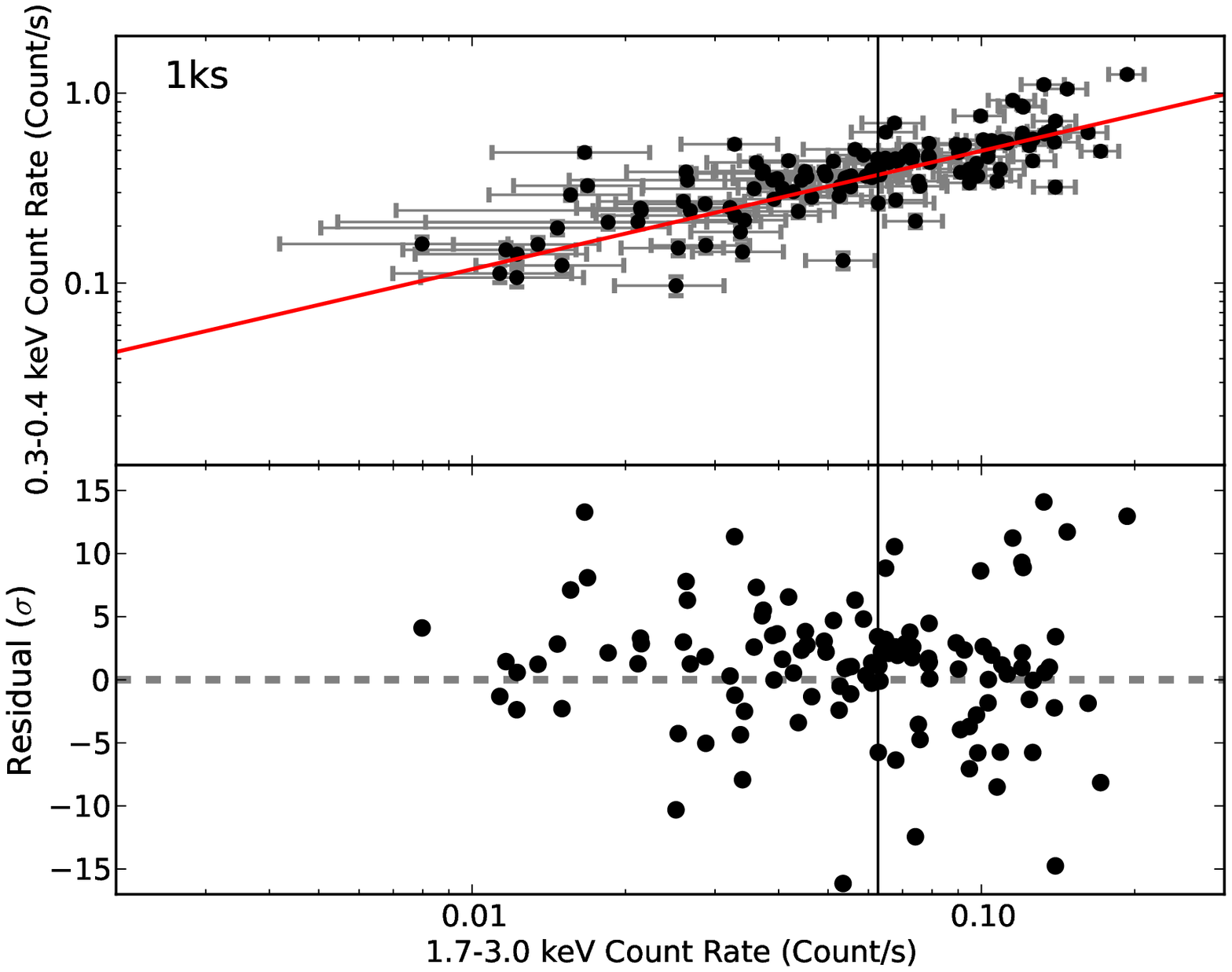}} 
\subfloat{\includegraphics[scale = 0.32]{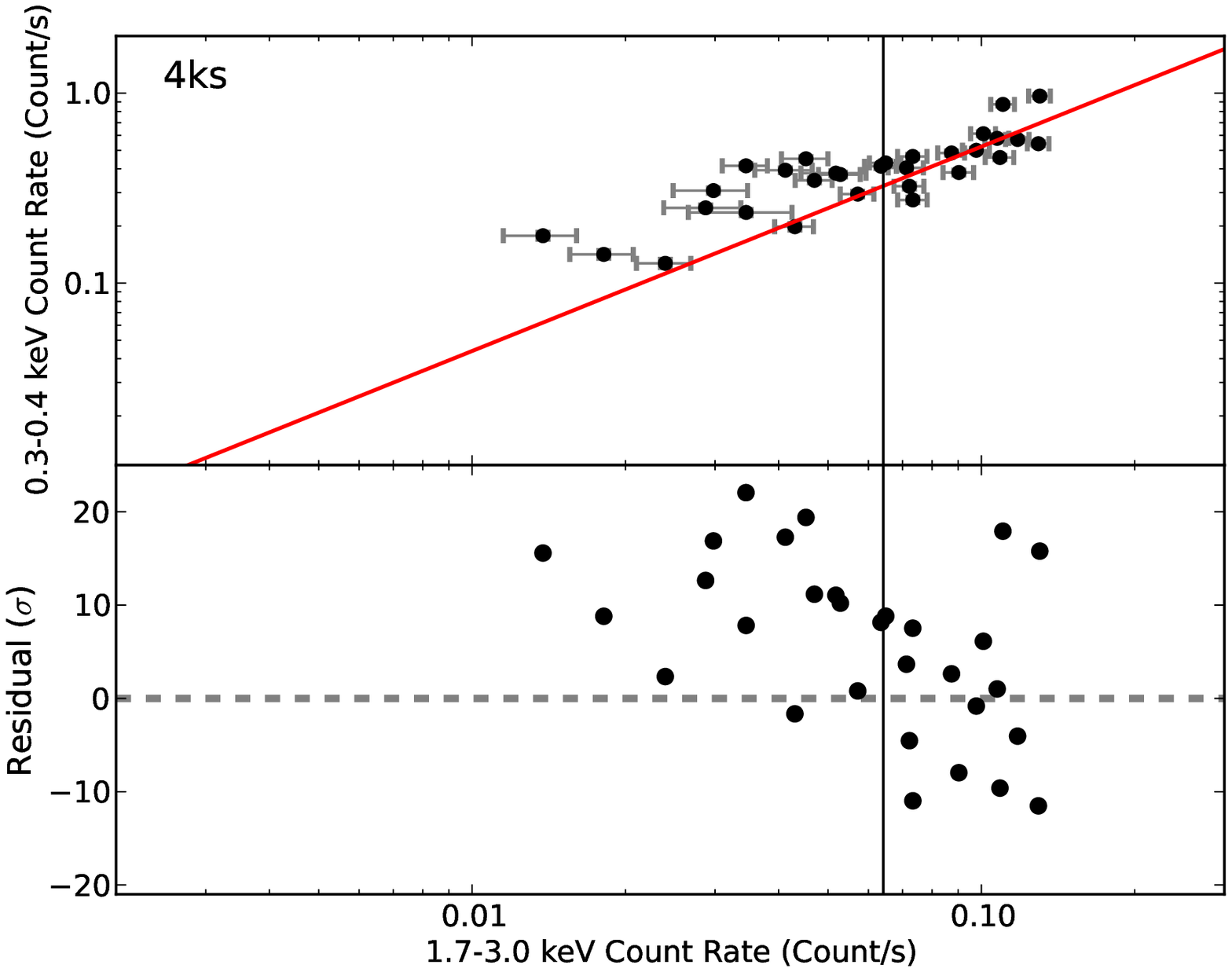}}
\subfloat{\includegraphics[scale = 0.32]{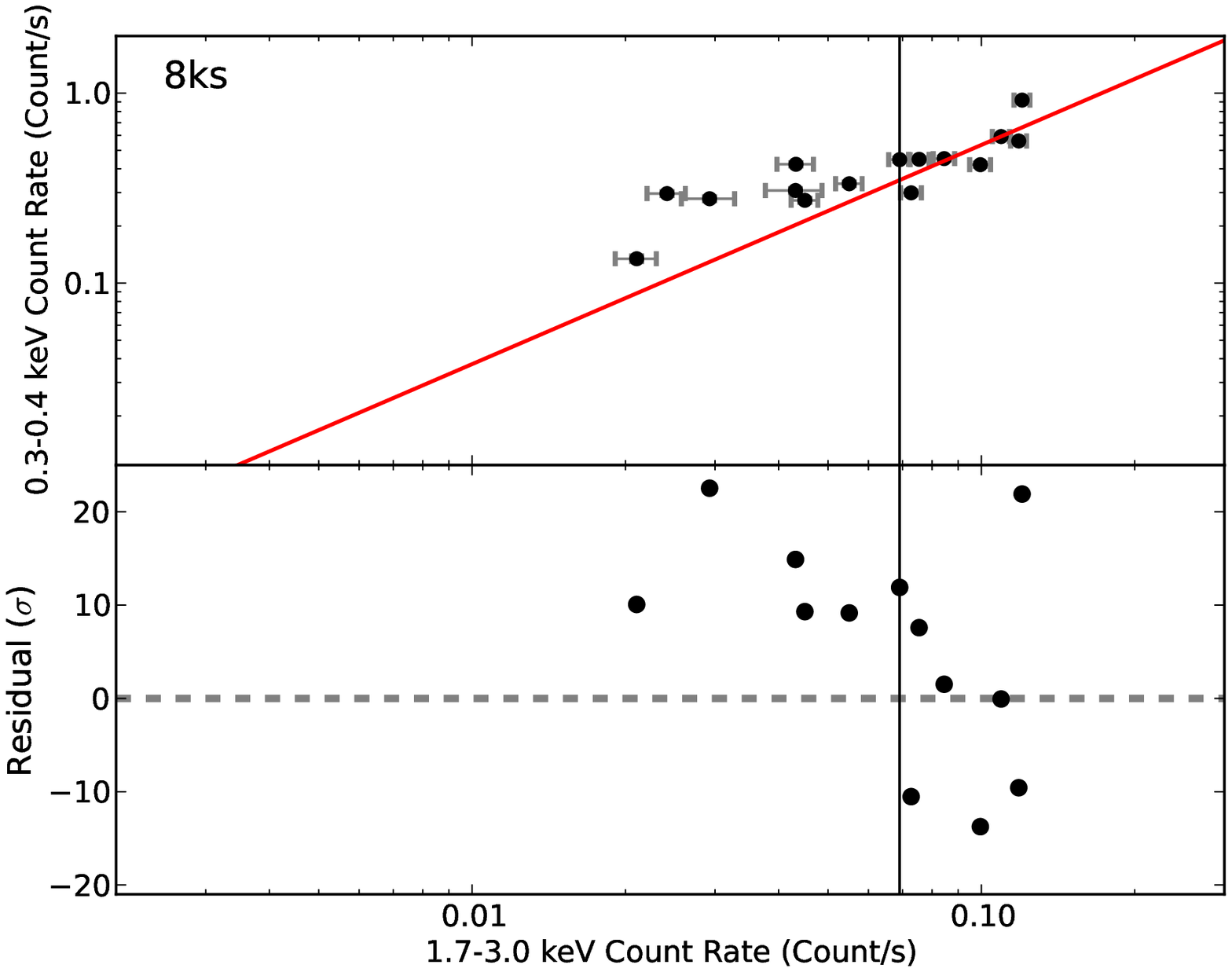}}\\
\subfloat{\includegraphics[scale = 0.32]{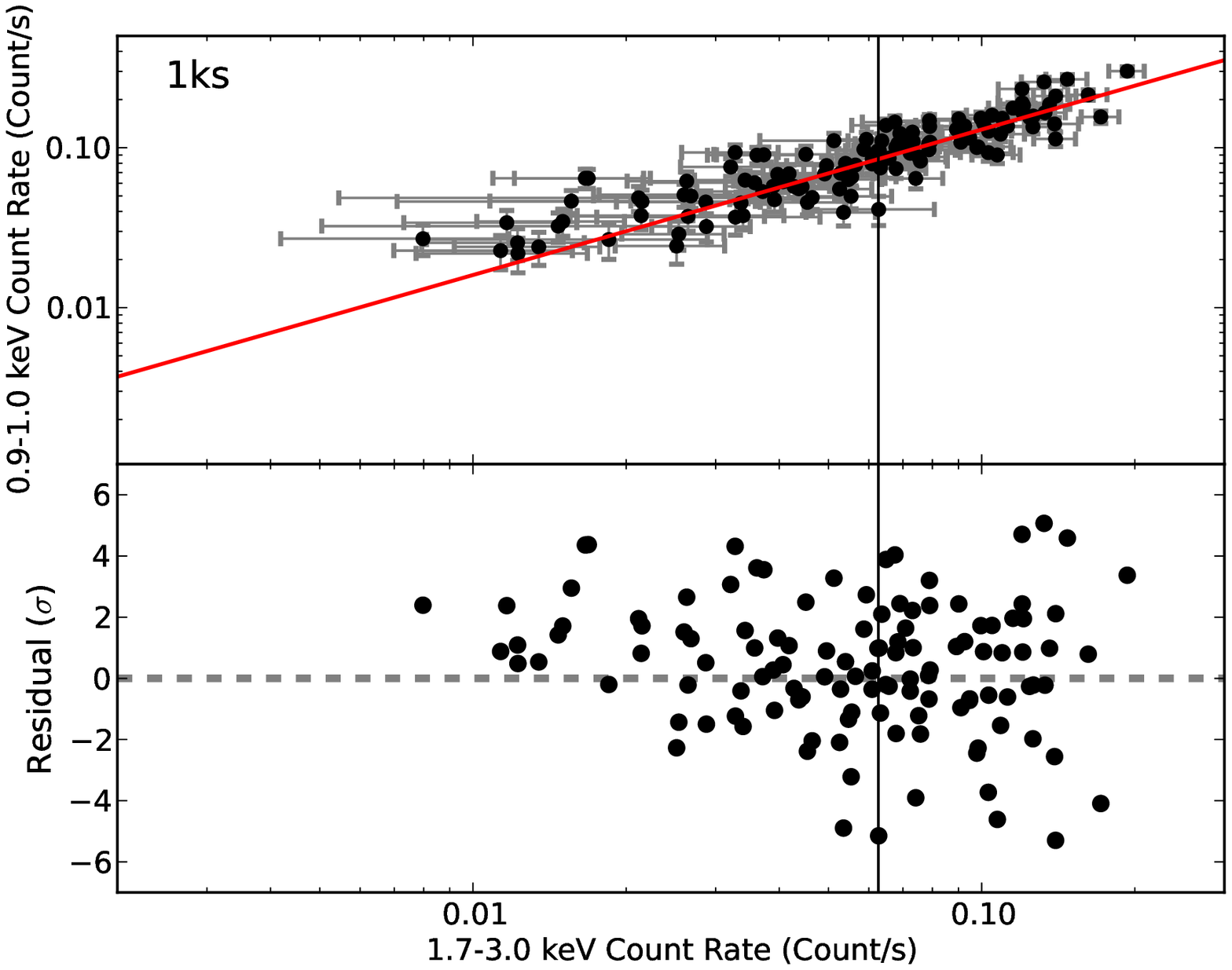}} 
\subfloat{\includegraphics[scale = 0.32]{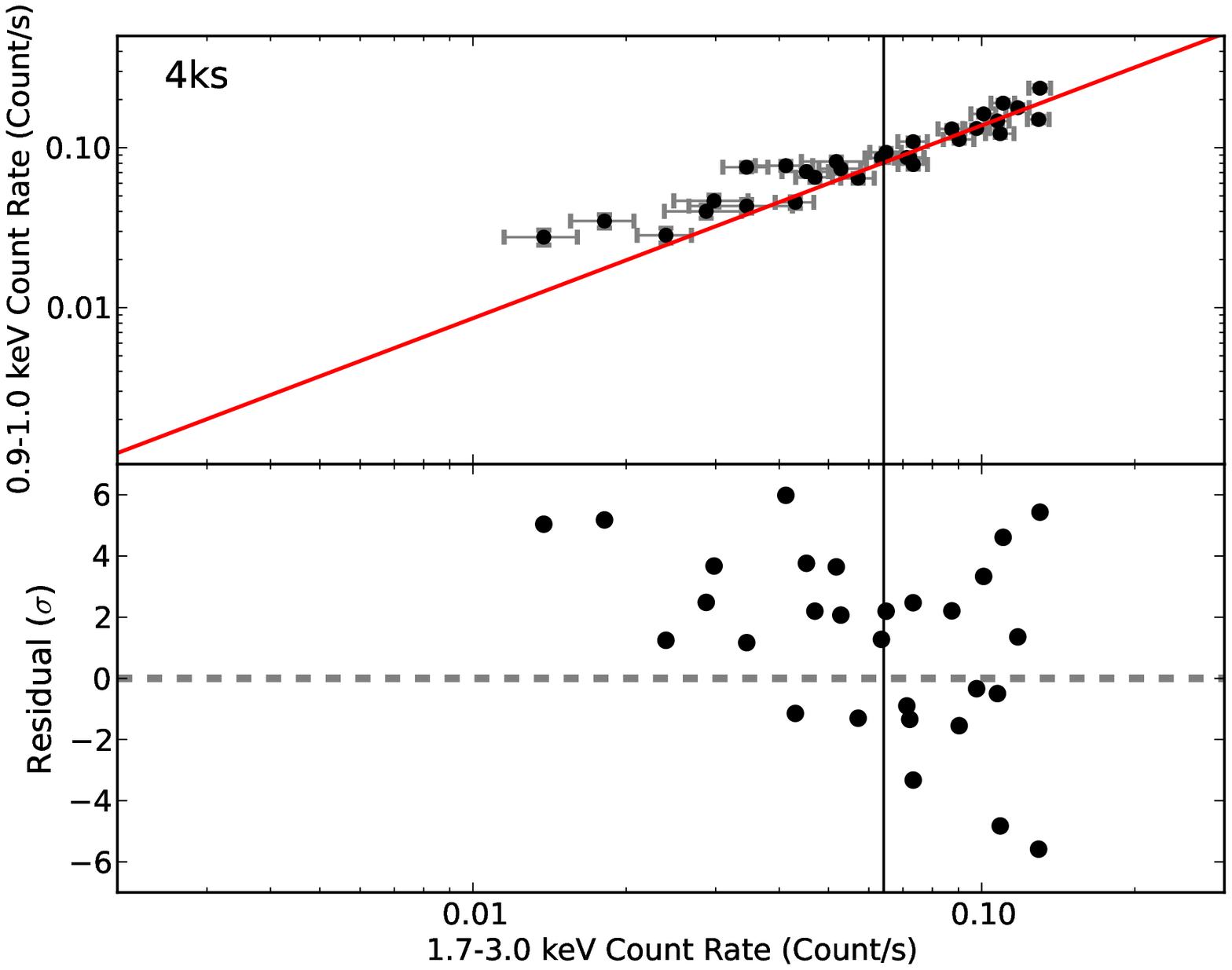}}
\subfloat{\includegraphics[scale = 0.32]{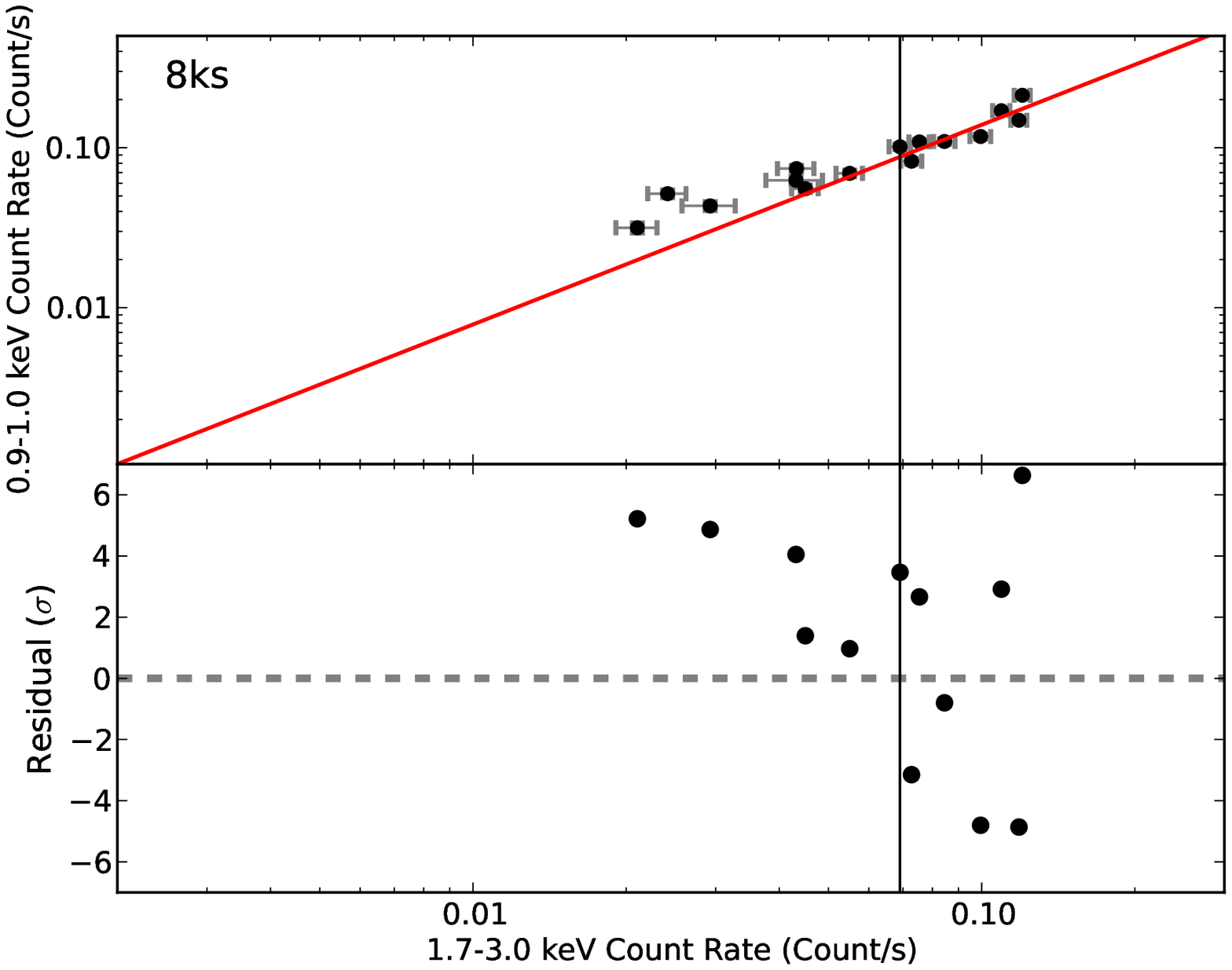}}\\
\subfloat{\includegraphics[scale = 0.32]{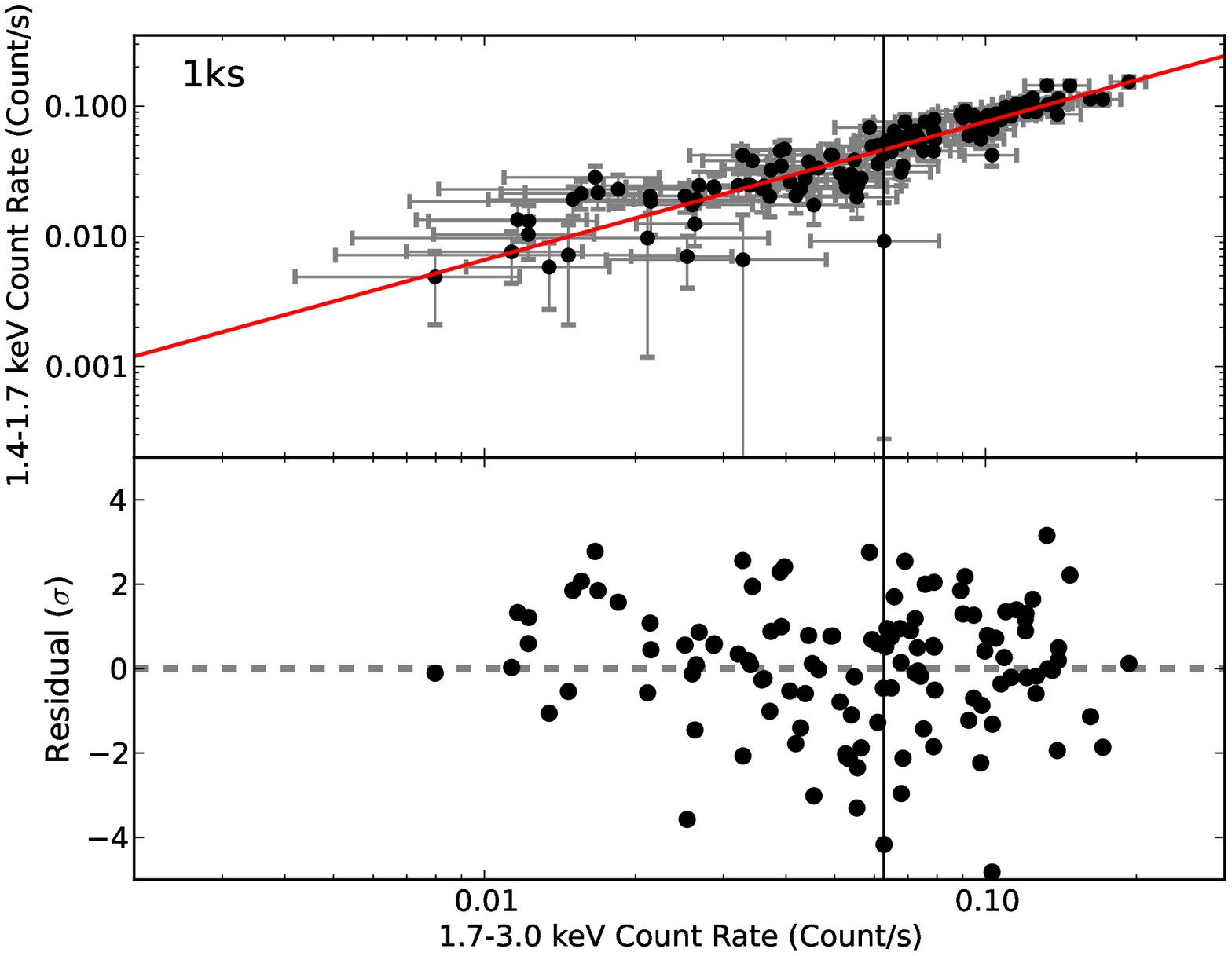}} 
\subfloat{\includegraphics[scale = 0.32]{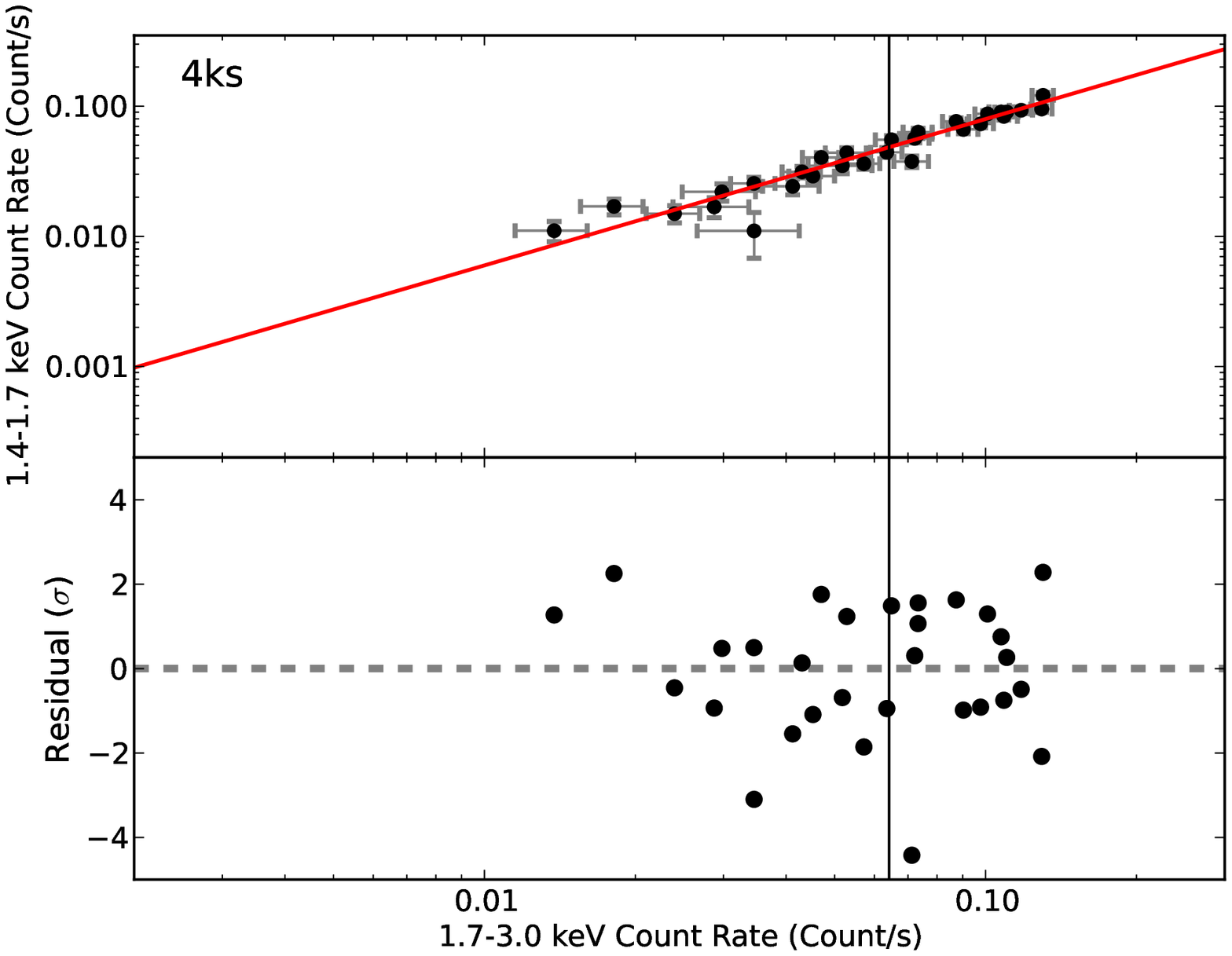}}
\subfloat{\includegraphics[scale = 0.32]{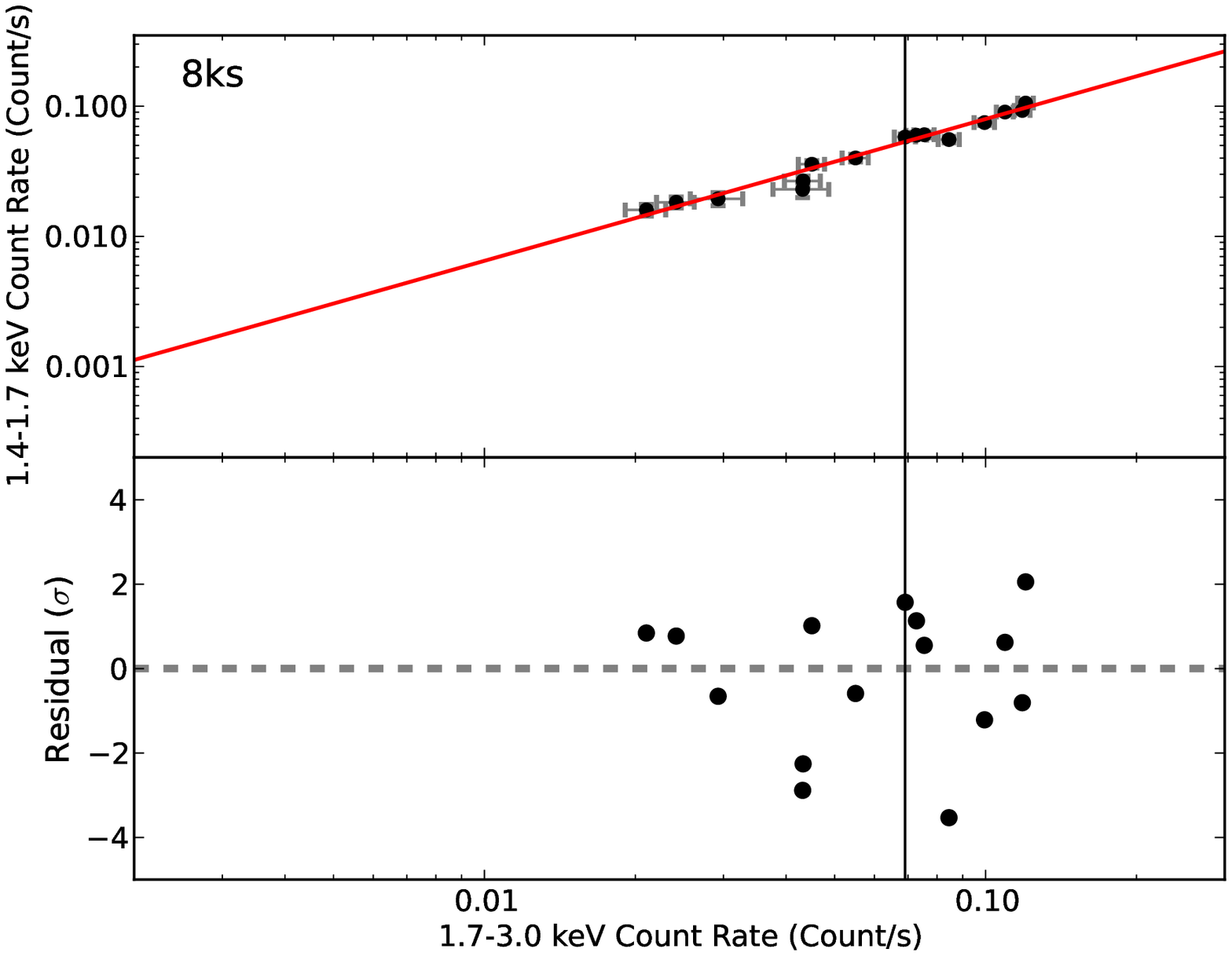}}
\caption{ 0.3--0.4, 0.9--1, and 1.4--1.7 vs 1.7--3\,keV flux--flux plots (top, middle, and bottom rows, respectively) for Obs. 3, obtained with $\Delta t_{\rm bin}=1, 4$, and 8\,ks (left, middle, and right columns, respectively). The solid red  line indicates the best-fit power-law relation to the `high-flux' data (see text for details). The vertical lines indicate the continuum band median. The best-fit residuals are plotted in the lower plan of each plot. Error bars are omitted in the residual plots for reasons of clarity .}
\label{fig:PLobs3}
\end{figure*}

\subsection{The flux--flux plots}
\label{subsec:fluxflux}

To construct the FFPs, we divide the $0.2-1.7$\,keV band into 11 energy sub-bands, with a width of 0.1\,keV from 0.2 up to 1\,keV. The width increases to 0.2\,keV for the next two sub-bands, and to 0.3\,keV in the final sub-band ($1.4-1.7$\,keV), so that the signal-to-noise ratio of the resulting light curves in this, and in the continuum bands, is larger than 3$\sigma$ (on average). 

We produced light curves in each energy band (including the continuum band) using a bin size, $\Delta t_{\rm bin}$, of 1, 4, and 8\,ks. We considered various $\Delta t_{\rm bin}$ values  to investigate whether $\Delta t_{\rm bin}$ affects the shape of the resulting FFPs or not. For the given data sets, a bin size smaller than 1\,ks will result in some data points having less than 10 counts in the $1.7-3\,{\rm\,keV}$ band. The error bars on these data points will be far from Gaussian, and, as a result, it would not be possible to fit the FFPs using traditional $\chi^2$ statistics. On the other hand, a $\Delta t_{\rm bin}$ larger than 8\,ks results in FFPs with a few points, and a reduced max/min dynamical range.

Figure\,\ref{fig:PLobs3} shows the 0.3--0.4, 0.9--1, and 1.4--1.7 vs 1.7--3\,keV FFPs (top, middle, and bottom rows, respectively) for Obs.\,3. They are representative of the FFPs of all observations. Left, middle, and right columns show the FFPs for $\Delta t_{\rm bin}=1$, 4, and 8\,ks, respectively. The soft and continuum band count rates are highly correlated, as is usual  for Seyferts. 

It is customary to fit the FFPs with linear relations in the form of: `soft band' flux = `hard/reference band' flux + constant. A stable component in the soft band is claimed when the best-fit intercept is significantly different from zero. We chose to follow a slightly different, and more general, approach in the modelling of the FFPs, and we fit them with a power-law (PL) relation instead.  If the underlying relation between soft and reference band counts is indeed linear, the best-fit model slope should be consistent with unity. If   there is a stable component, whose flux is smaller than the primary flux, we would expect a flattening of the FFP at low reference band fluxes and the detection of a non-zero constant, just like when we fit the data with a linear function. 

\subsection{The effects of the light curve bin size}
\label{subsec:timebin}

To investigate the dependence of the FFPs on $\Delta t_{\rm bin}$, we first considered the `high-flux' data points in these plots, i.e. the points with a count rate in the continuum band that is higher than its median (the vertical lines in Fig.\,\ref{fig:PLobs3} indicate the median in the case of the Obs.\,3 plots). In this way, we avoid complications that may be associated with a possible flattening at low frequencies, resulting from the presence of a stable spectral component. 

We fitted the high-flux part of the FFPs with a power-law (PL) relation of the form
 $$y = \alpha x^\beta ,$$
where $y$ and $x$ represent the `soft' and the continuum band count rates, respectively. The model normalisation, $\alpha$, and slope, $\beta$, were both left to vary freely during the fit. The fits were performed using the Python routine {\tt MPFIT}\footnote{\url{https://code.google.com/p/astrolibpy/source/browse/mpfit/mpfit.py}}, based on the Fortran routine {\tt LMFIT} \citep{Mor78}, taking into account the errors on the $y$-axis (i.e. the errors of only the soft energy band counts). The solid red lines in Fig.\,\ref{fig:PLobs3} indicate the best-fit PL lines, extrapolated to lower fluxes as well. The bottom panels in each plot show the best-fit residuals (i.e. the ratio (data-model)/$\sigma$).  The residuals at the high-flux part of the plots are uniformly scattered around zero, and there is no indication of any systematic residual trend. 

\begin{figure}[!ht]
\centering
\includegraphics[scale=0.45]{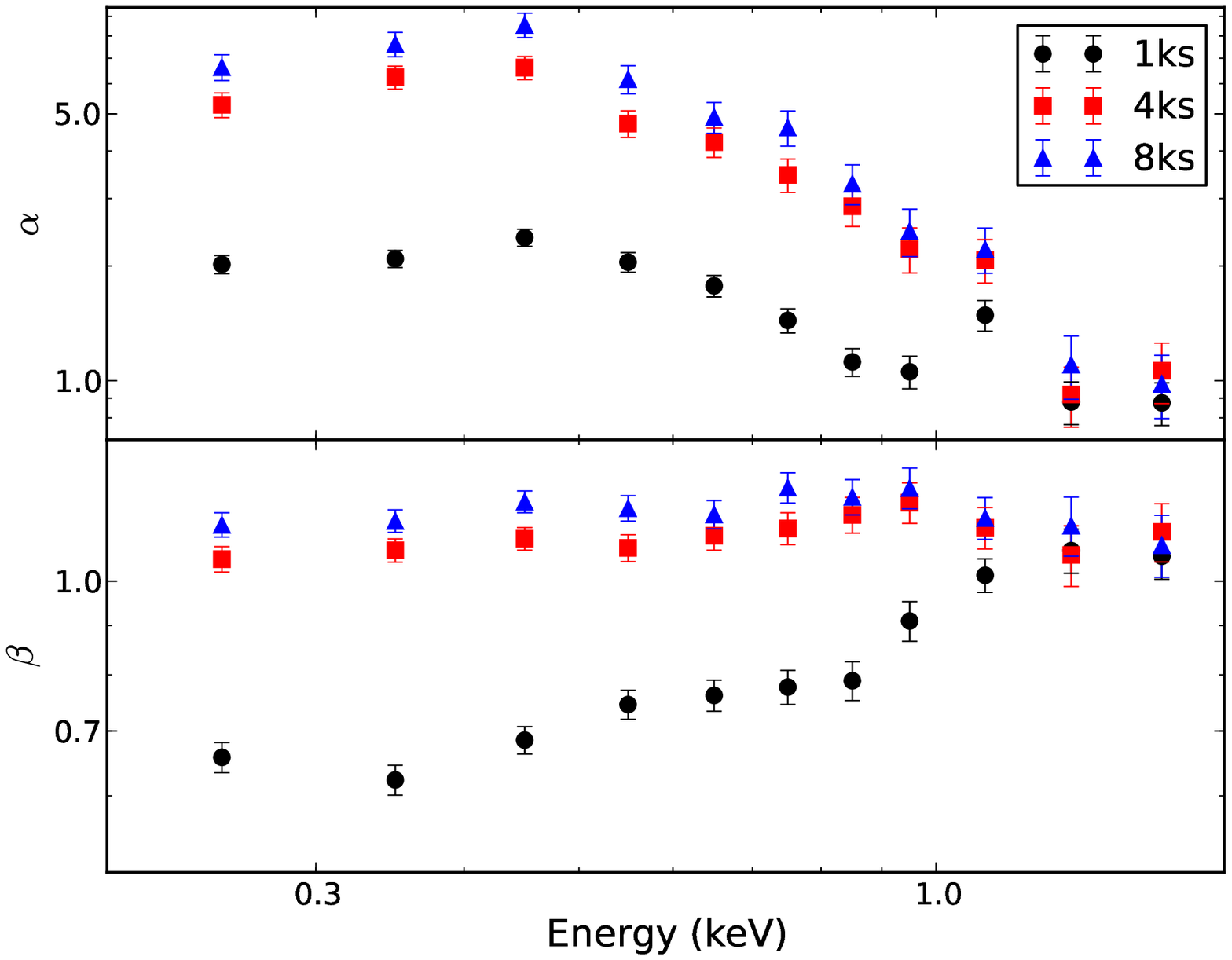}
\caption{Best-fit $\alpha$ and $\beta$ values (top and bottom panels, respectively), obtained by fitting a PL model to the `high-flux' data points (see text for details) of the 1, 4, and 8\,ks FFPs of Obs.\,3 (black filled circles, red filled squares, and blue filled triangles, respectively)}
\label{fig:PLparam-obs3}
\end{figure}

Figure\,\ref{fig:PLparam-obs3} shows the best-fit PL normalisation and slope values for all the FFPs of Obs.\,3 (top and bottom panel, respectively). Black, red, and blue points indicate the results for the 1, 4, and 8\,ks binned light curves, respectively. The best-fit $\alpha$ and $\beta$ values of the 4 and 8\,ks binned light curves are consistent with each other. This is not the case with the 1\,ks  binned data: the best-fit $\alpha_{\rm 1ks}$ values are systematically smaller than the respective 4 and 8\,ks values at energies below $\sim 1$\,keV (where the soft excess is more pronounced), and the same is true for the best-fit slopes as well.  

This can be seen from the plots in the two upper rows in Fig.\,\ref{fig:PLobs3}. The $x-$ and $y-$axis range is the same in all panels. It is clear that the 4 and 8\,ks best-fit models (solid red lines in the middle and right-hand panels) cannot fit well the high-flux part of the 1\,ks flux--flux plots (shown in the left panels). We observe similar differences between the best-fit ($\alpha_{1 ks},\beta_{1 ks}$) and ($\alpha_{4/8 ks},\beta_{4/8 ks}$) values in the other observations as well. These differences strongly suggest that the choice of the time bin size does affect the shape of the resulting FFPs. We provide a possible explanation for the effect of light curve bin size on the FFPs, below.

The 4 and 8\,ks FFPs are not a binned version of the 1\,ks FFPs. The data binning has been performed in the time, and not in the flux, domain. A large $\Delta t_{\rm bin}$ value may result in a FFP which is distorted and does not correspond to the intrinsic plot, if the intrinsic relation between the count rates in two energy bands is non linear (i.e. $\beta\neq 1$), and the source is variable. For example, supposing that the intrinsic relation between the `soft' and `hard' band counts, C$_{\rm sb}$ and C$_{\rm hb}$, respectively, is in the form of  C$_{\rm sb}=a{\rm C}_{\rm HB}^{\beta_{\rm intr}}$, and that both bands are variable. The light curve binning implies a binning of the source signal in the time domain in the form

$${\rm C}_{\rm bin}(t)=\frac{1}{\Delta t_{\rm bin}}\int_{t}^{t+\Delta t_{\rm bin}}{\rm C}(t)dt\, ,$$ 
where ${\rm C}_{\rm bin}$ stands for `binned counts'. If $\beta_{\rm intr}\neq 1$, then since

$$\int_{t}^{t+\Delta t_{\rm bin}}{\rm C}_{\rm HB}^{\beta_{\rm intr}}(t)dt\neq \left(\int_{t}^{t+\Delta t_{\rm bin}}{\rm C}_{\rm HB}(t)dt\right)^{\beta_{\rm intr}},$$ 
the relation between C$_{\rm HB,bin}(t)$ and C$_{\rm SB,bin}(t)$ will not be identical to the intrinsic one.

IRAS 13224--3809 is a highly variable source in X-rays. \cite{Gon12} have estimated its 0.2--2\,keV power spectrum. It has a power-law like shape, with a normalization of $\sim 10^{-5}$ and a slope of $\sim -2$. This implies that the fractional root mean square variability amplitude on timescales of 8, 4, and 1\,ks is 28\%, 20\%, and 10\%, respectively\footnote{By definition, the fractional mean square variability amplitude over a given time period, say $T$, is equal to the integral of the power-spectrum density from $1/T$ up to infinity.}. So, when we bin the light curves using a bin size of 8 and 4\,ks, we typically average data points which scatter around the mean by a factor of $\sim 0.3$ and $0.2$, respectively. The average scatter should only be  $\sim 10\%$ in the 1\,ks binned light curves. Perhaps then, the $\Delta\beta$ differences we observe between the 1\,ks and the 4/8 \,ks flux--flux plots are due to the large variability amplitude of the source and the intrinsically non-linear flux--flux relations in this source. Henceforth, we have decided to study only the 1\,ks binned flux--flux plots  in this work.

\begin{figure*}
\centering
\subfloat{\includegraphics[scale = 0.31]{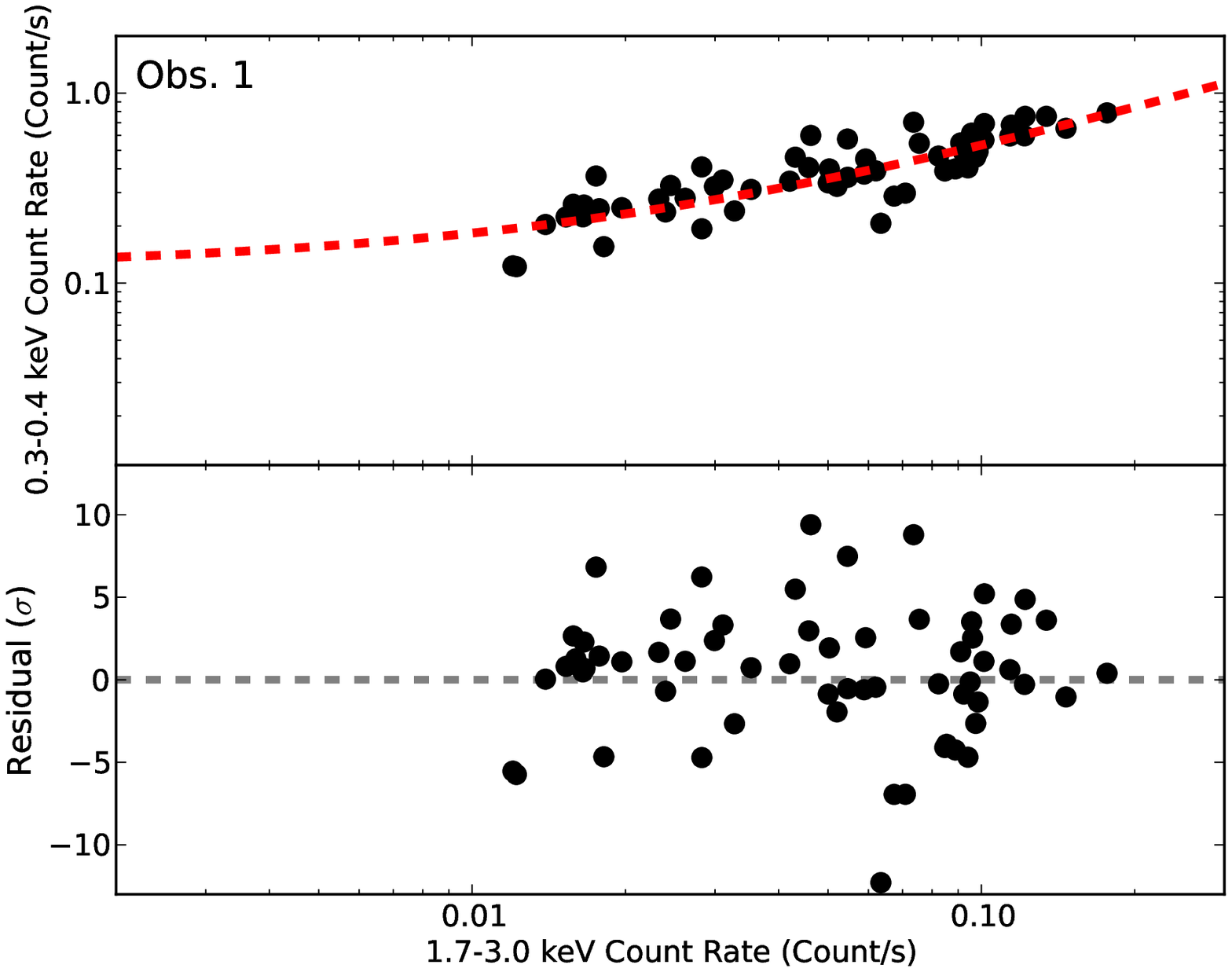}} 
\subfloat{\includegraphics[scale = 0.31]{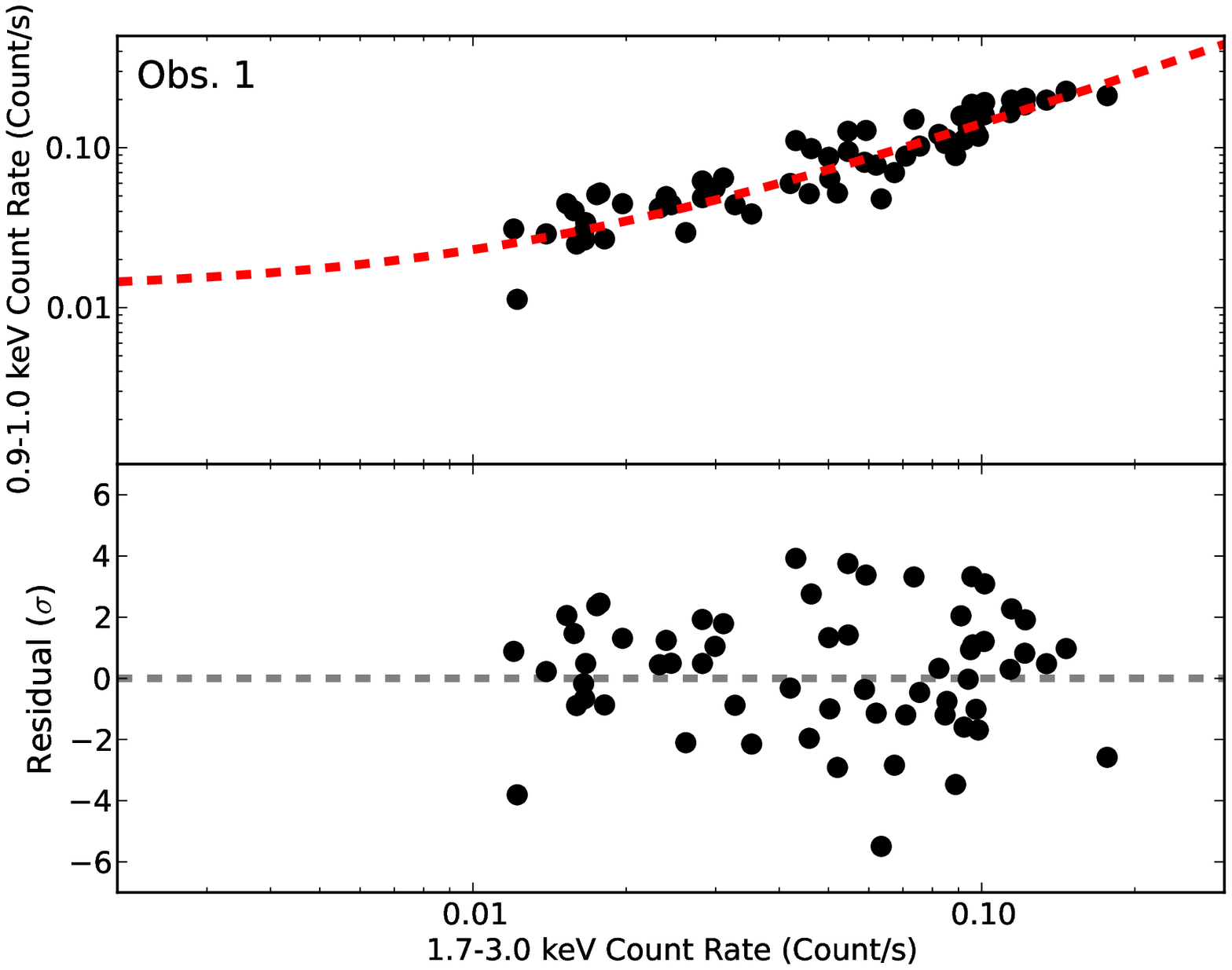}}
\subfloat{\includegraphics[scale = 0.31]{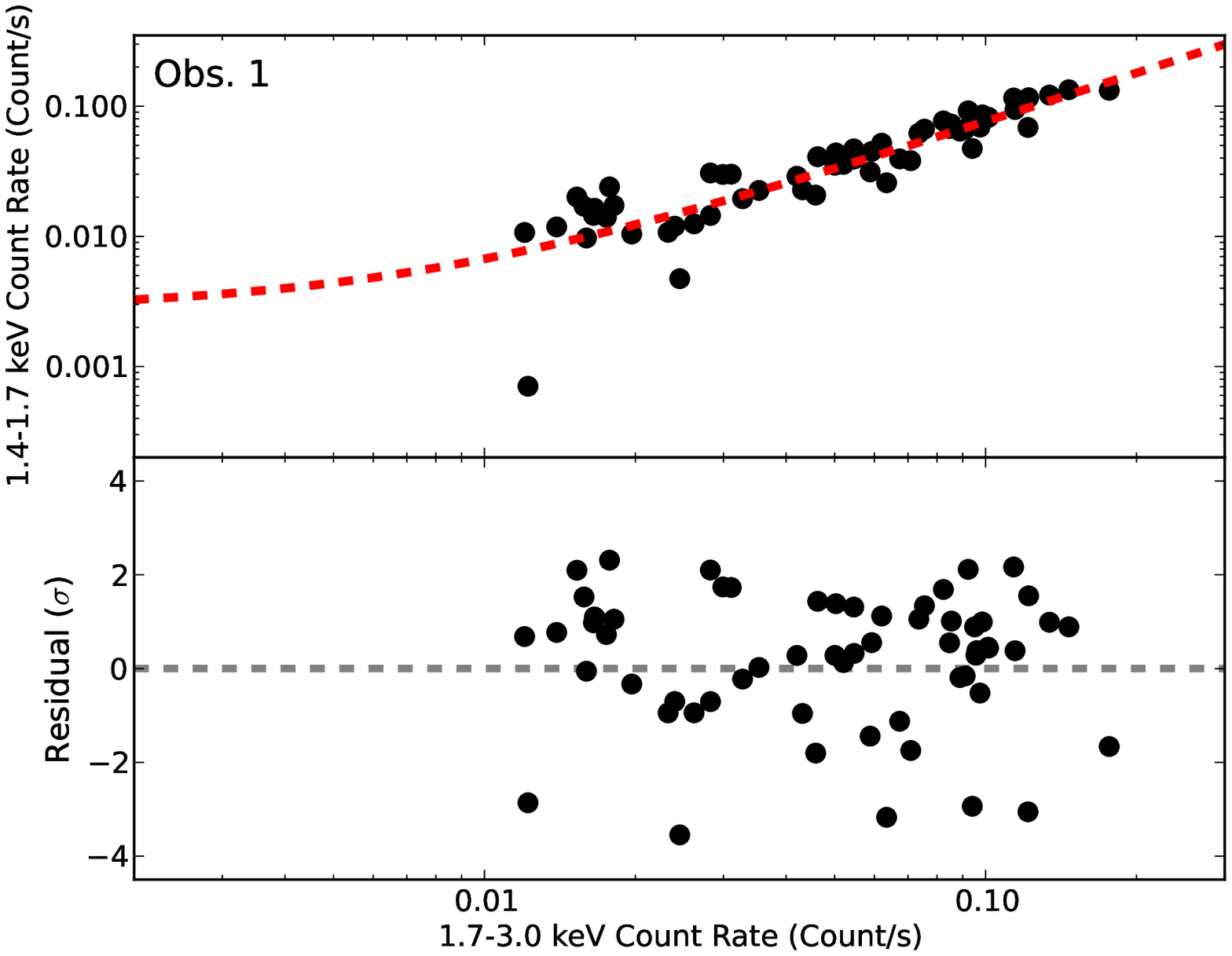}}\\
\subfloat{\includegraphics[scale = 0.31]{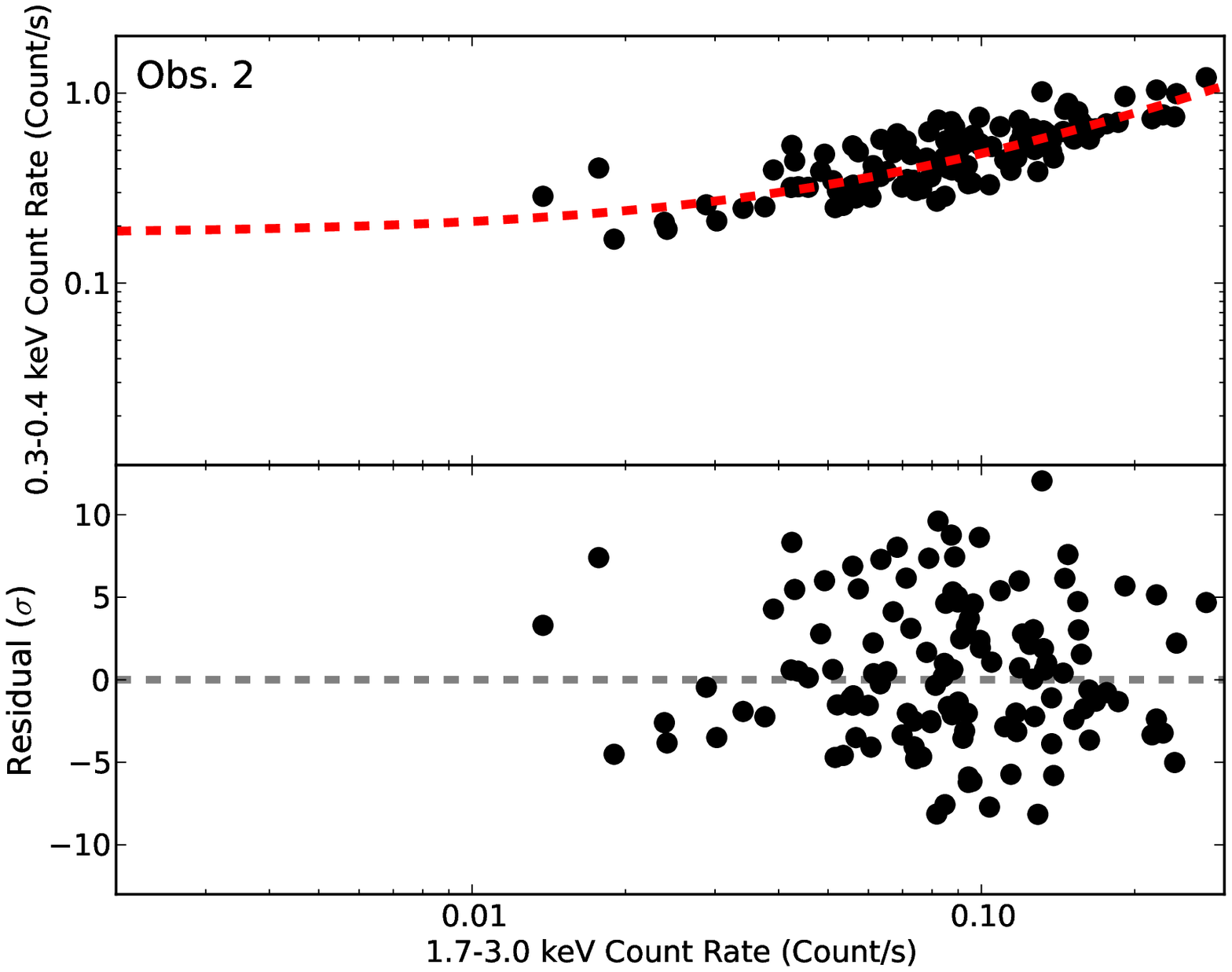}} 
\subfloat{\includegraphics[scale = 0.31]{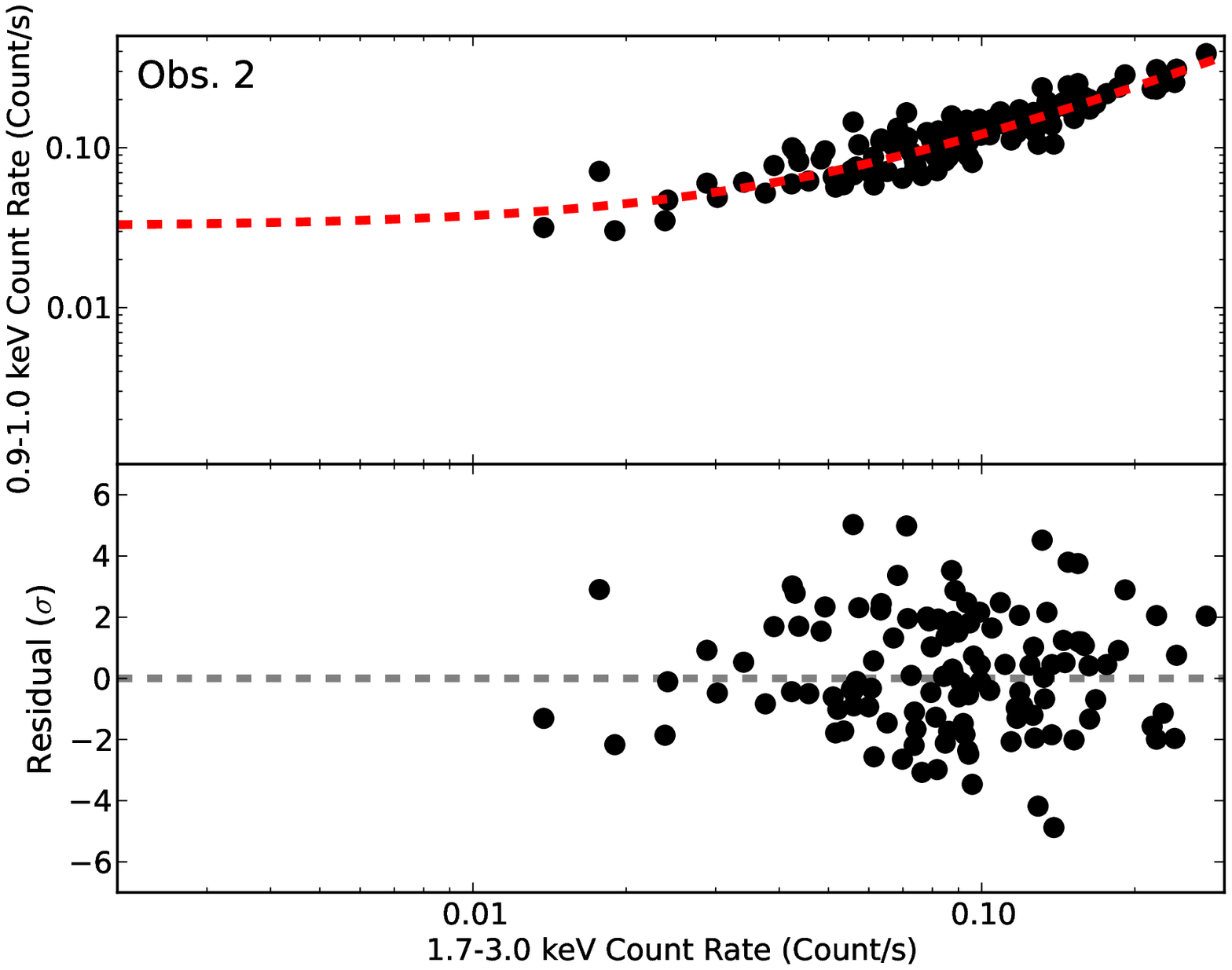}}
\subfloat{\includegraphics[scale = 0.31]{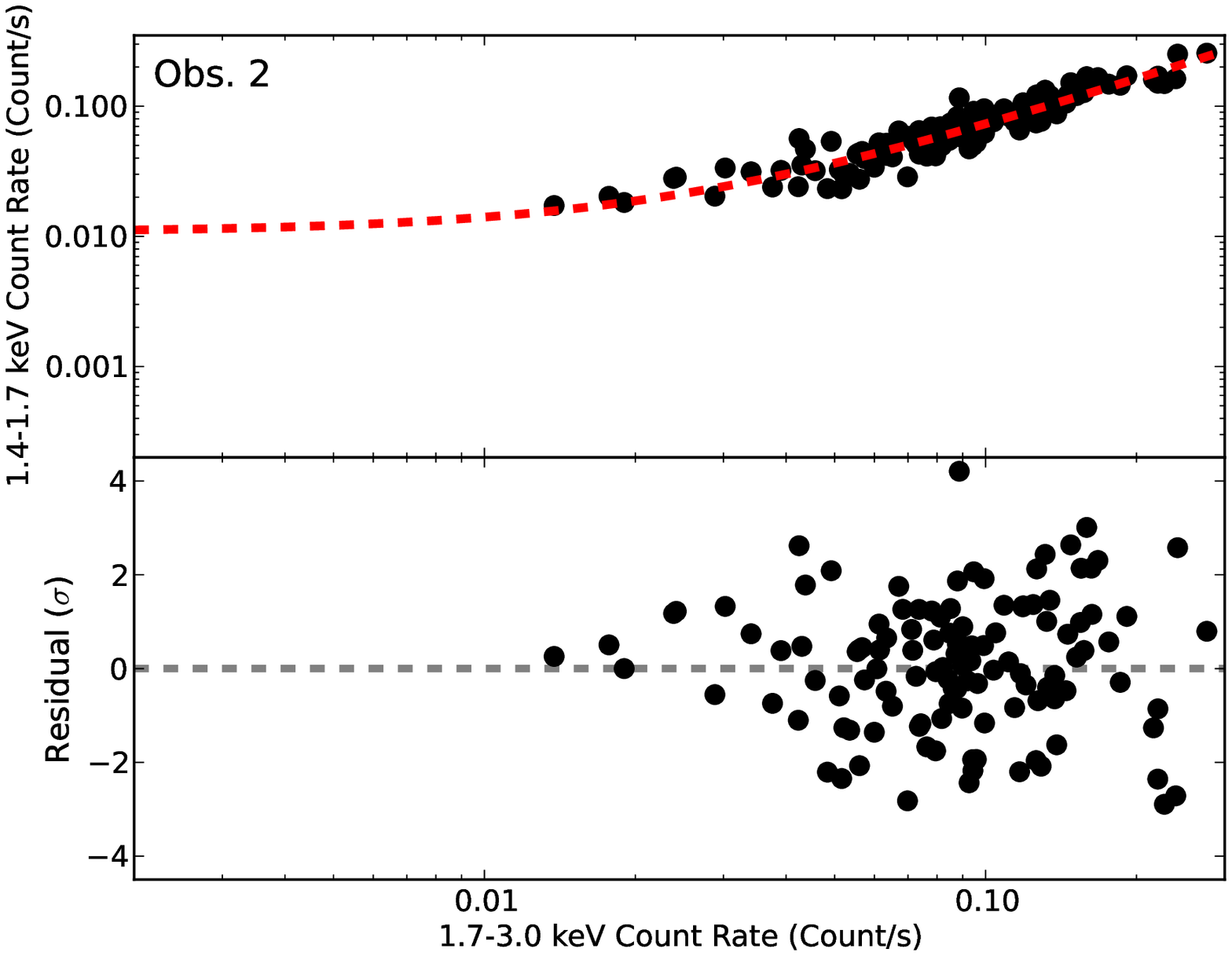}}\\
\subfloat{\includegraphics[scale = 0.31]{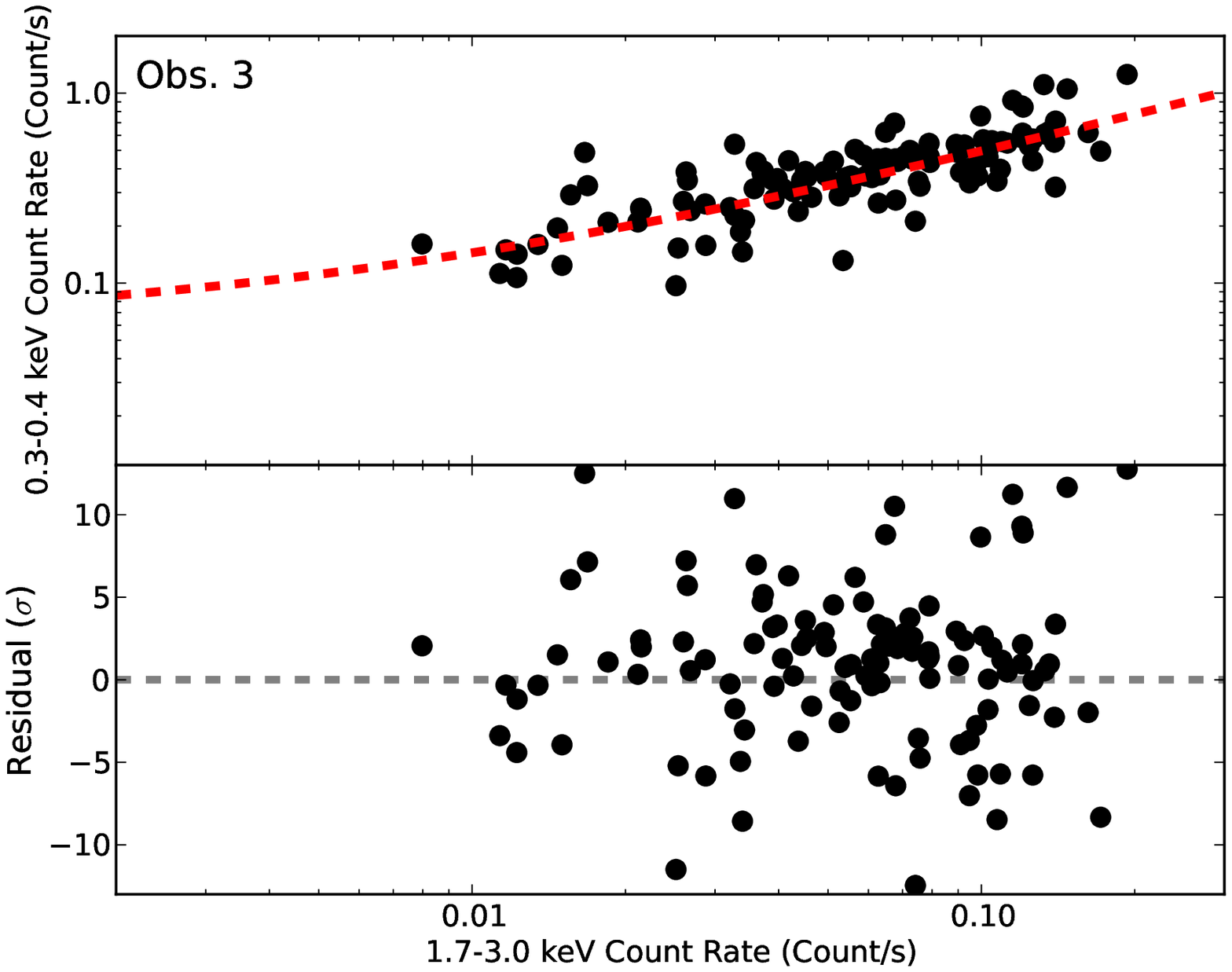}} 
\subfloat{\includegraphics[scale = 0.31]{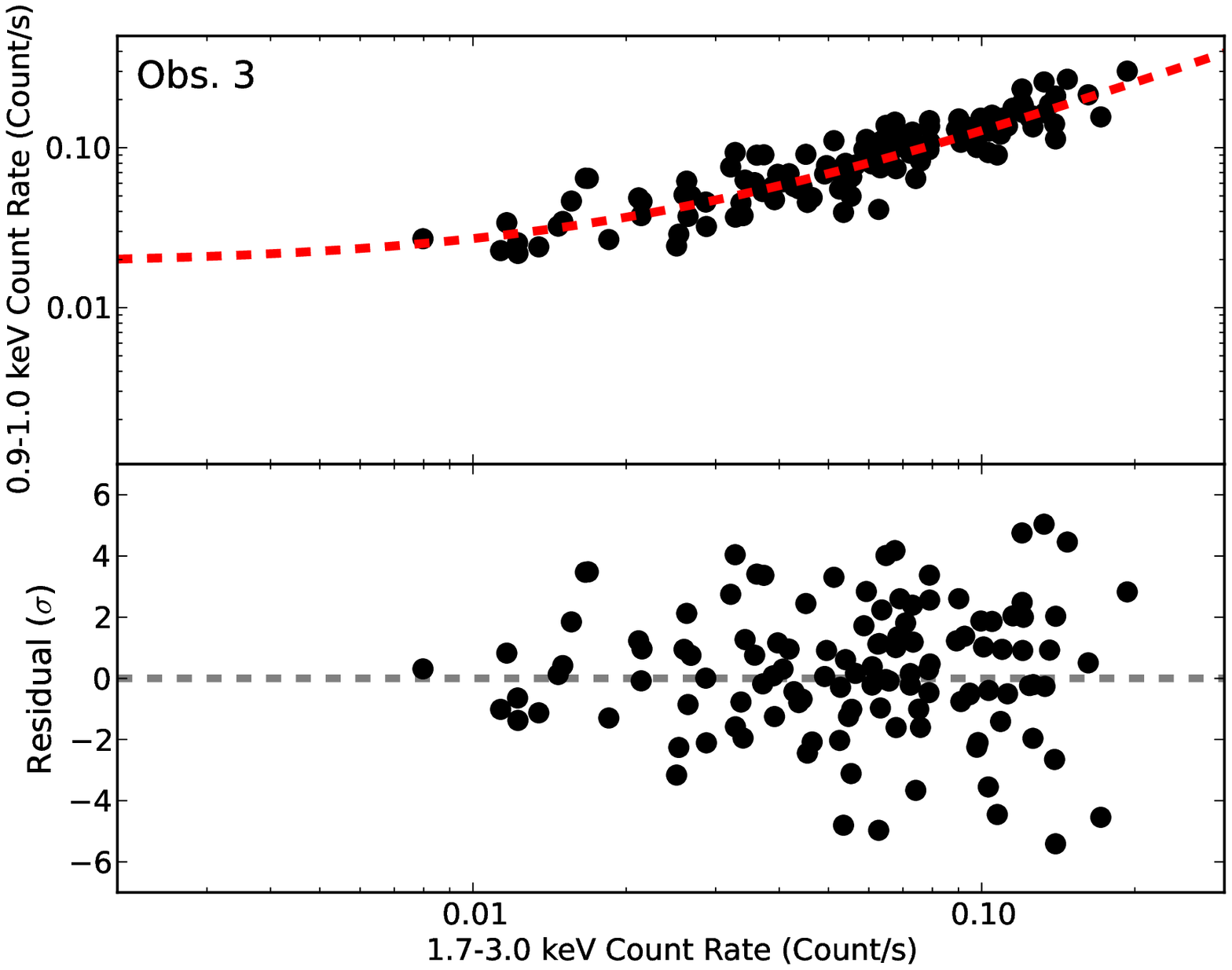}}
\subfloat{\includegraphics[scale = 0.31]{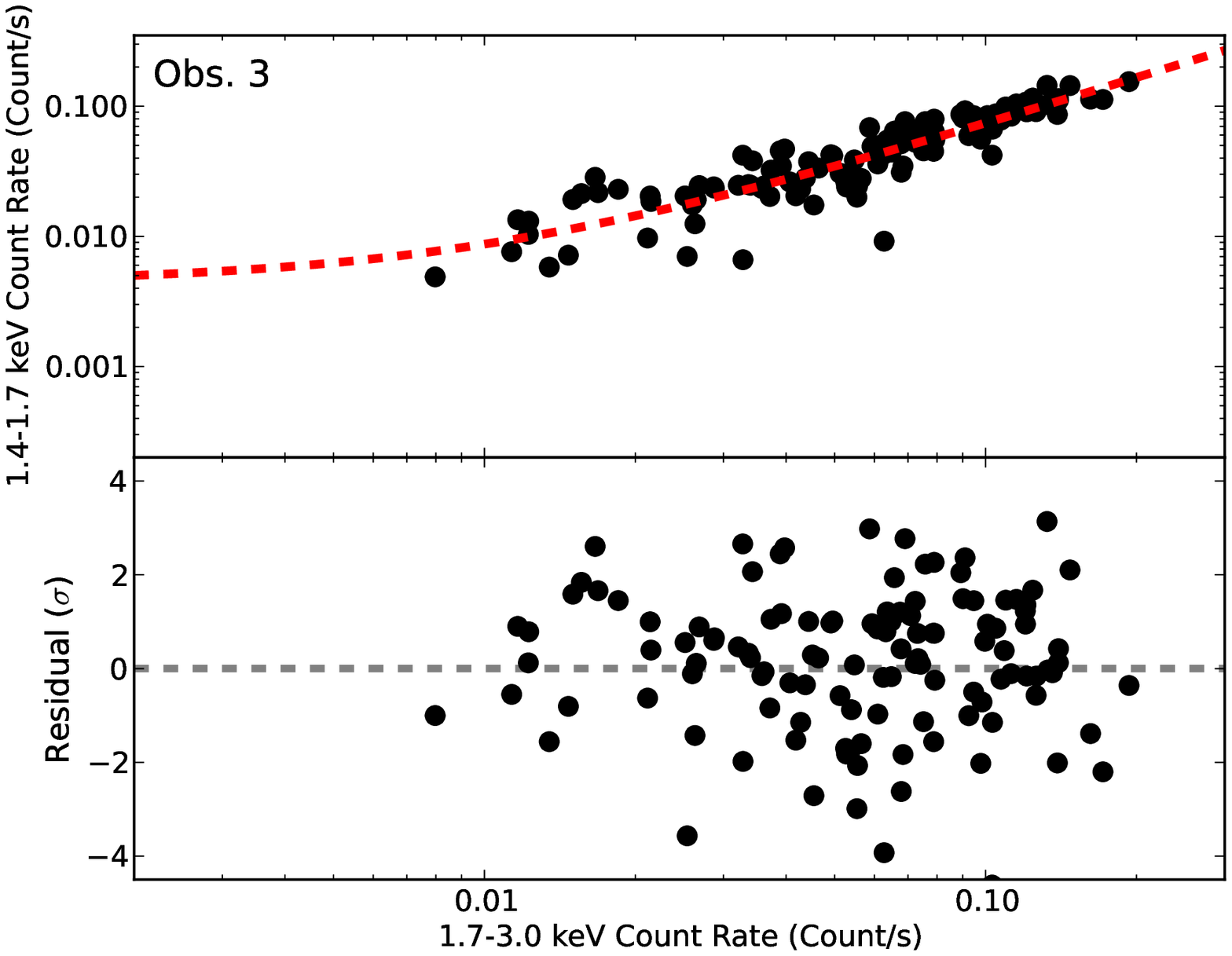}}\\
\subfloat{\includegraphics[scale = 0.31]{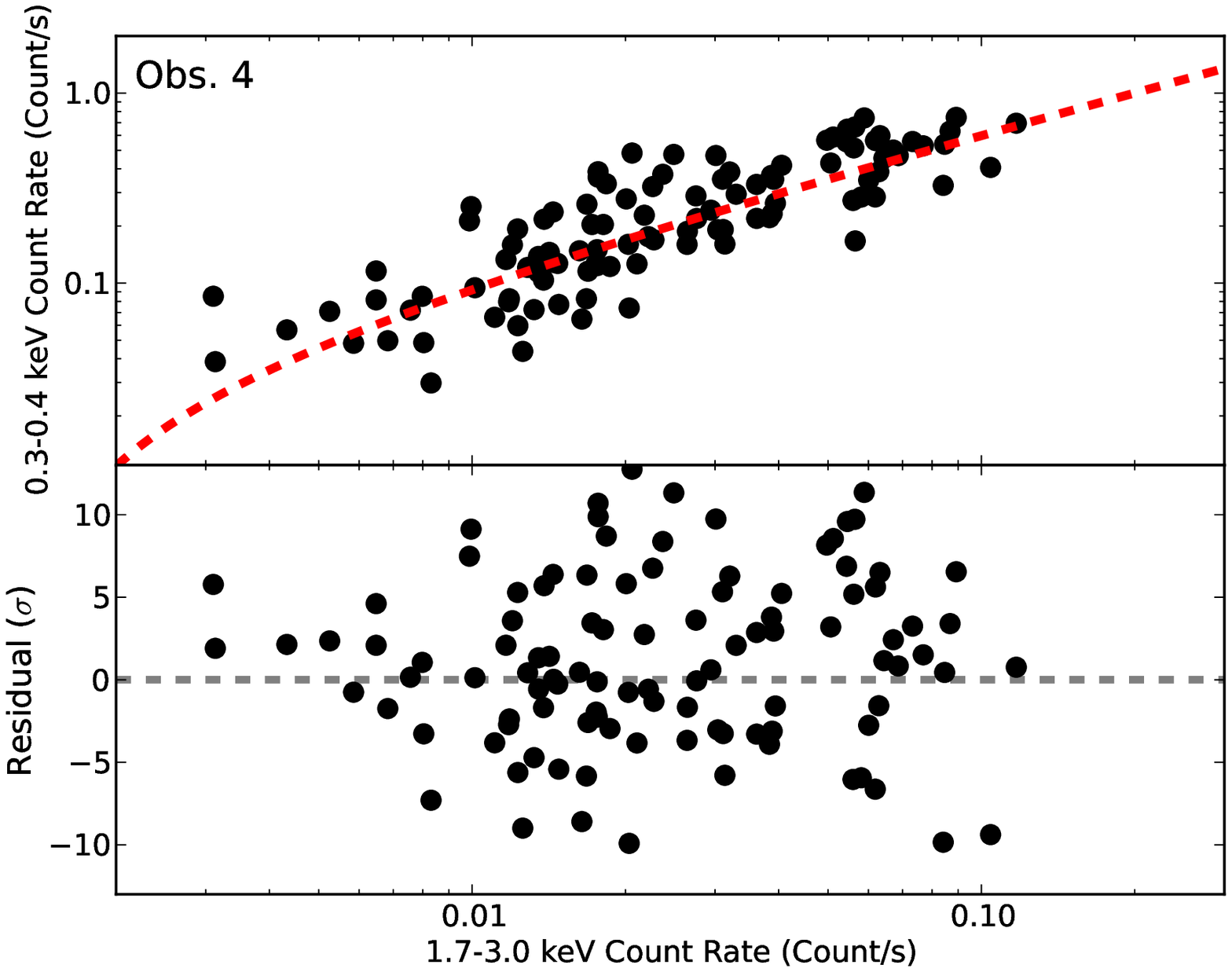}} 
\subfloat{\includegraphics[scale = 0.31]{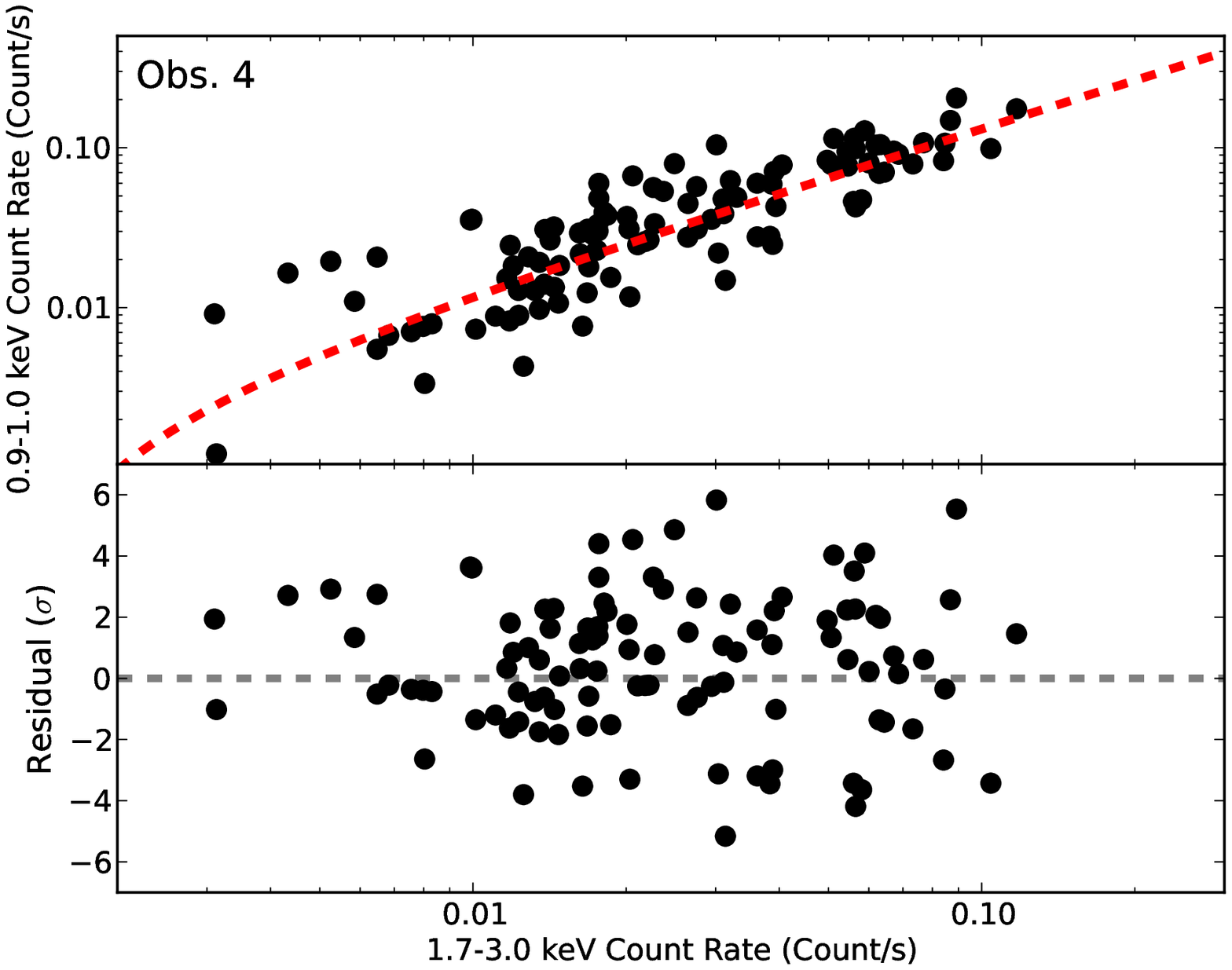}}
\subfloat{\includegraphics[scale = 0.31]{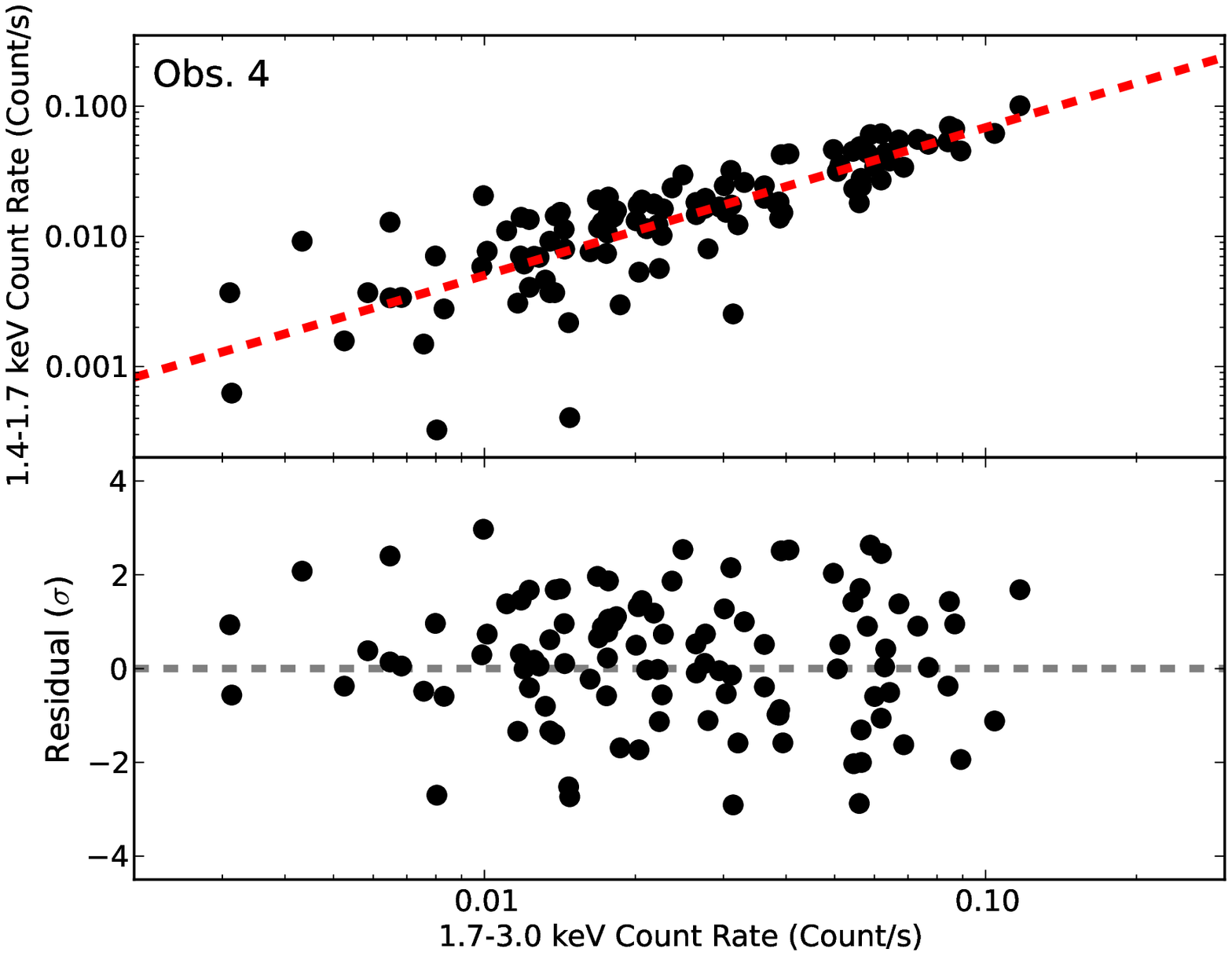}}\\
\subfloat{\includegraphics[scale = 0.31]{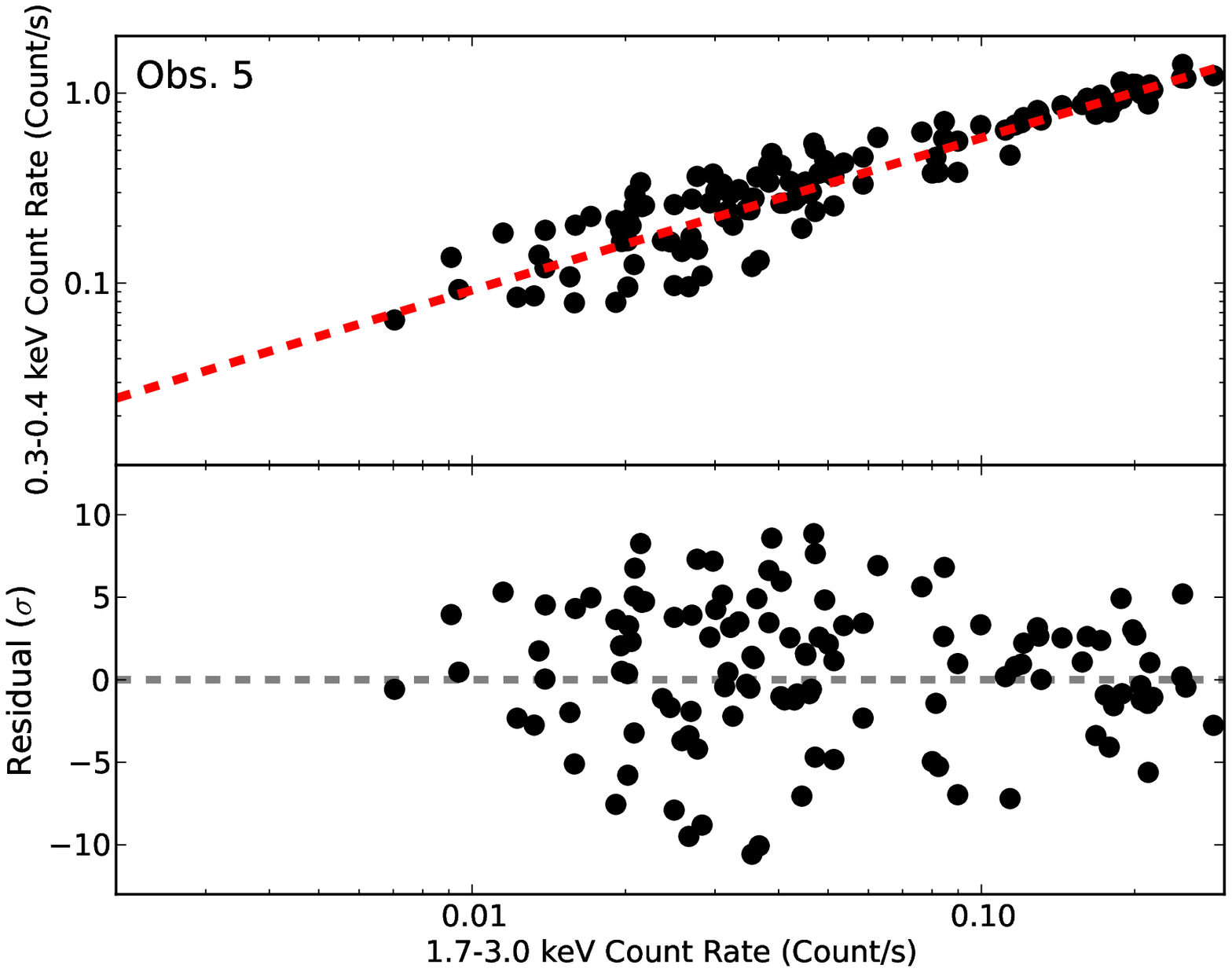}} 
\subfloat{\includegraphics[scale = 0.31]{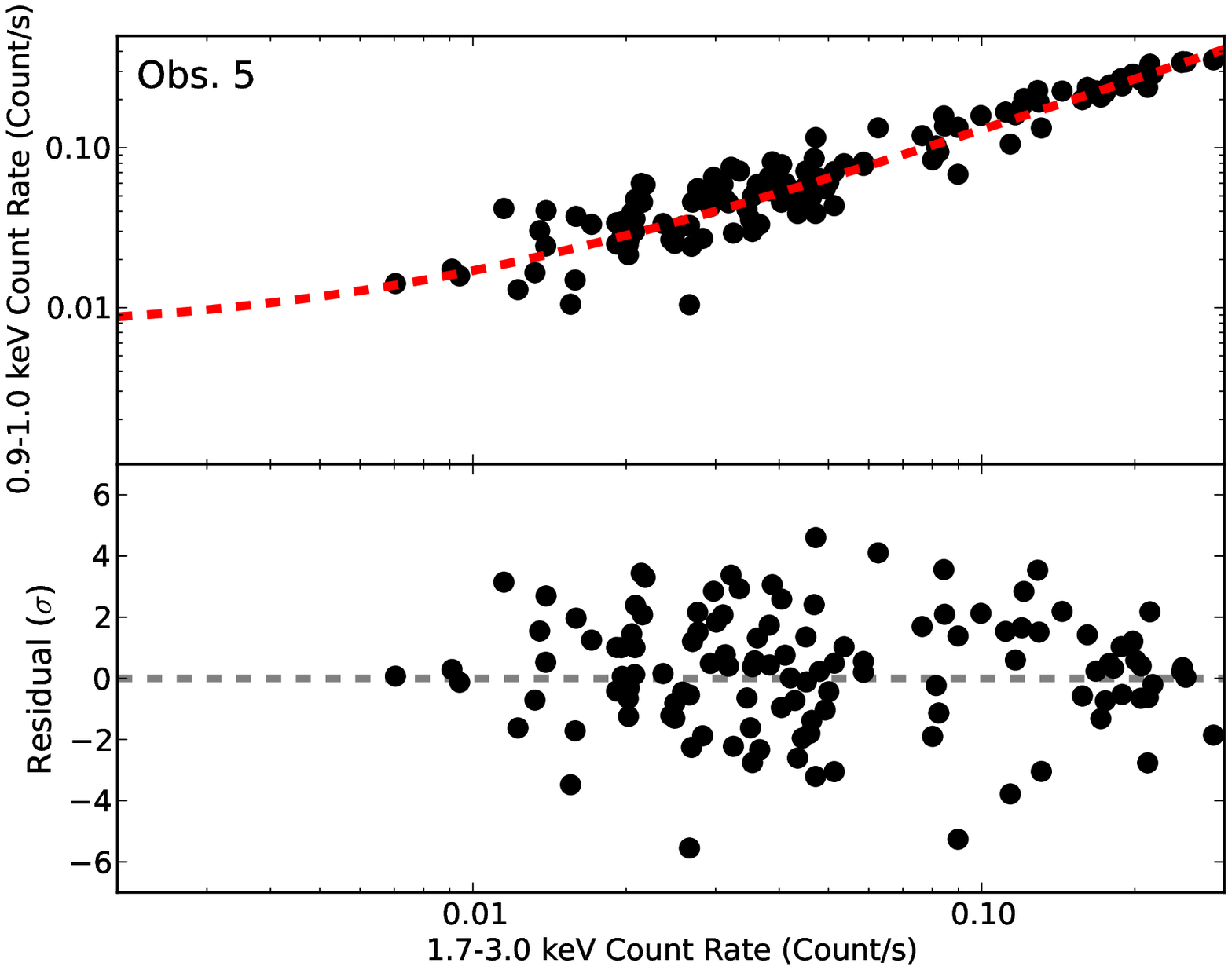}}
\subfloat{\includegraphics[scale = 0.31]{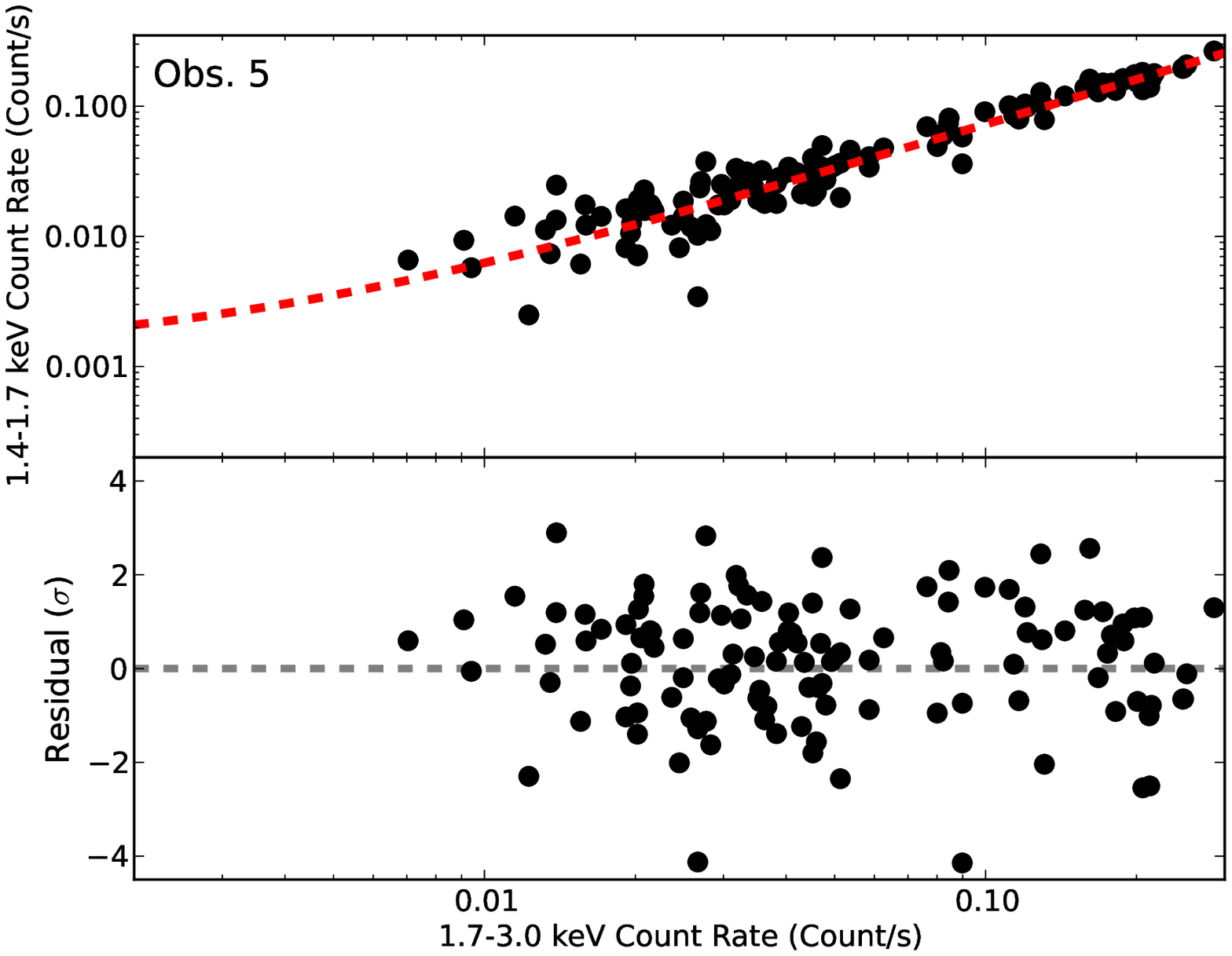}}
\caption{0.3--0.4, 0.9--1, and 1.4--1.7 vs 1.7--3\,keV FFPs (left, middle, and right columns, respectively) for all observations obtained for a bin size of 1\,ks. The dashed red  line indicates the best-fit PLc relation. The best-fit residuals are plotted in the lower plan of each plot. The errors are similar to those plotted in Fig.\,\ref{fig:PLobs3} but were not plotted for clarity reasons.}
\label{fig:PLcfit}
\end{figure*}

\subsection{The case of a soft-band `constant component'}
\label{subsec:constExc}

The residual's plots in the left-hand column of Fig.\,\ref{fig:PLobs3} show a data `excess' above the interpolation of the best-fit PL to low count rates.  The excess is more pronounced in the  top and middle panels (i.e. in the 0.3--0.4 and 0.9--1.0 vs 1.7--3\,keV plots), where the soft excess is stronger in this source (see Fig.\,\ref{fig:ratio}). As we have argued above, such a flattening in the FFPs at low count rates could be due to the presence of a constant soft-band component. To quantify its presence, spectral shape, and strength,  we fitted all the FFPs with a power-law plus constant (PLc) model in the form of
$$ y = \alpha_\text{PLc} x^{\beta_\text{PLc}}+ c. $$
We fitted the data using {\tt MPFIT}. All parameters were left free during the fitting.  

The best-fit results are listed in Table\,\ref{table:PLcparam}. The lines within each observation list the results for the 0.3--0.4, 0.4--0.5, 0.5--0.6, 0.6--0.7, 0.7--0.8, 0.9--1, 1--1.2, 1.2.--1.4, and 1.4-1--1.7 vs 1.7--3\,keV band FFPs. Figure\,\ref{fig:PLcfit} shows the 1\,ks, 0.3--0.4, 0.9--1, and 1.4--1.7 vs 1.7--3\,keV band flux--flux plots (left, middle, and right columns, respectively) for all observations. The dashed lines show the best-fit PLc models. The bottom panels in the  plots show the best-fit residuals, which are uniformly scattered around the best-fit models, with no indication of any systematic discrepancies. The PLc model  describes  the overall trend in the FFPs well, and the large $\chi^2$ reduced values are due to low-amplitude, but significant, soft-band variations that are not related to the primary source flux. 

The last column in the same table lists the root mean square deviation, $\sigma_{\rm rms}$, of the data from the model. This is estimated by

$$ \sigma_{\rm rms} = \left( \frac{1}{N}\sum_i \frac{ (y_i - y_{{\rm model},i})^2 - \sigma_i^2}{y_{\text{model},i}^2}\right)^{1/2}, $$
where N is the number of data points in the FFP, $y_i$ and $\sigma_i$ represents the observed count rate in the soft energy band and its corresponding error, and $y_{{\rm model},i}$ represents the model count rate in the same band. In practice, $\sigma_{\rm rms}$ measures the average discrepancy between the best-fit model and the data points, also taking  the data errors into account (but only of the dependent variable). The average data-to-model deviations are $\sim 20-35$\% for all FFPs of all observations, except Obs.\, 4, where the deviations increase to almost $\sim 50$\% (max).

\begin{figure}
\centering
\includegraphics[scale=0.5]{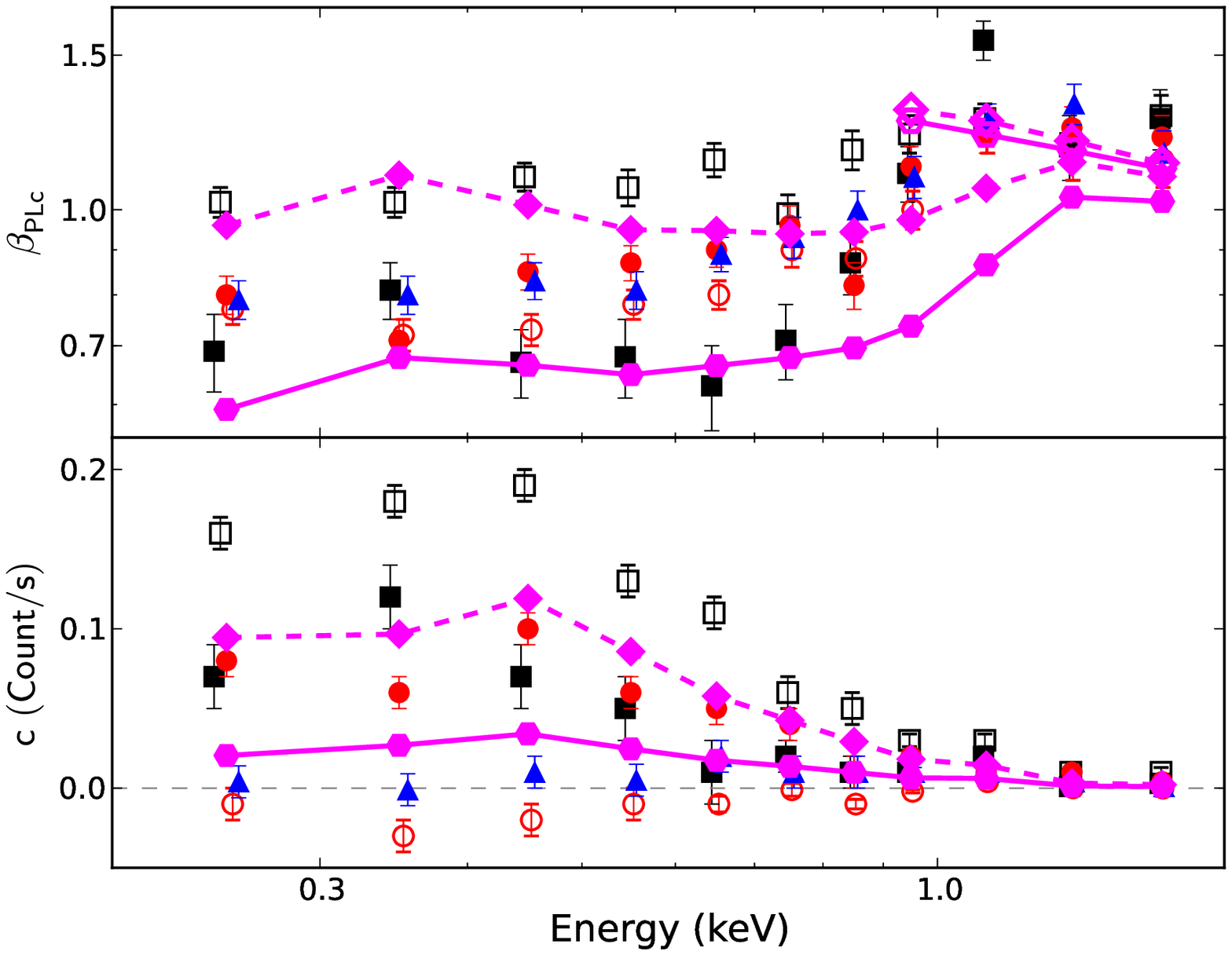}
\caption{Best-fit parameters $\beta_{\rm PLc}$ (top) and $c$ (bottom), obtained by fitting a PLc model to the 1-ksec binned light curves for Obs.\,1 (black filled squares), Obs.\,2 (black open squares), Obs.\,3 (red filled circles), Obs.\,4 (red open circles), and Obs.\,5 (blue filled triangles).The model parameters, obtained by fitting a PLc relation to simulated FFPs, are also plotted for the high- and low-flux case models we considered (magenta-filled diamonds and magenta-filled hexagons, respectively). The model parameters, in the case of a variable primary in flux and spectral slope without adding an excess component, are also plotted for the high-flux and low-flux cases (open magenta diamonds and hexagons, respectively; see Sec.\,\ref{subsec:simulation} for details).}
\label{fig:PLc-Param}
\end{figure}

Figure\,\ref{fig:PLc-Param} shows a plot of the best-fit $\beta_{\rm PLc}$ and $c$ values (top and bottom panels, respectively) for all observations.  The best-fit slopes are flatter than 1 at energies below 1\,keV, where the soft excess is stronger. The only exception is the data from Obs.\,2, where the PL slopes are close to unity. The best-fit slopes increase towards unity with increasing energy until $\sim 0.9-1$ keV, and then are steeper than 1 at higher energies. 

The best-fit $c$ values are consistent with zero in all the FFPs of Obs.\, 5. This is the case with the FFPs of Obs.\, 4 as well. In fact, most of the best-fit $c$ values at energies below 1\,keV are negative for this observation.  On the other hand, the best-fit $c$s   are  significantly different from zero at energies below $\sim 1$\,keV in the FFPs of Obs.\,2 and Obs.\, 3. They are positive in the case of the Obs.\,1 FFPs as well, although in many cases at a level lower than $3\sigma$.

\begin{table}
\centering
\caption{Best-fit PLc models to the FFPs of all observations. }
\begin{tabular}{ c c c c c}
\hline \hline
 $\alpha_{\rm PLc}$ & $\beta_{\rm PLc}$ & $c_{\rm PLc}$ & $\chi^2$ & $\sigma_{\rm rms}$  \\ 
  & & & & \\ \hline 
                                        &                               &       Obs.\,1                 &       dof=58  &                       \\      
                1.9     $\pm$   0.2     &       0.69    $\pm$   0.07    &       0.07    $\pm$   0.02    &       859.8   &       0.24            \\      
                2.7     $\pm$   0.3     &       0.81    $\pm$   0.06    &       0.12    $\pm$   0.02    &       973.0   &       0.25            \\      
                2.2     $\pm$   0.2     &       0.67    $\pm$   0.06    &       0.07    $\pm$   0.02    &       1028.7  &       0.25            \\      
                1.7     $\pm$   0.2     &       0.68    $\pm$   0.07    &       0.05    $\pm$   0.02    &       880.1   &       0.26            \\      
                1.5     $\pm$   0.2     &       0.63    $\pm$   0.07    &       0.01    $\pm$   0.02    &       591.5   &       0.24            \\      
                1.3     $\pm$   0.2     &       0.71    $\pm$   0.07    &       0.02    $\pm$   0.01    &       602.6   &       0.26            \\      
                1.5     $\pm$   0.2     &       0.87    $\pm$   0.07    &       0.01    $\pm$   0.01    &       378.3   &       0.27            \\      
                1.6     $\pm$   0.3     &       1.10    $\pm$   0.08    &       0.01    $\pm$   0.004   &       241.7   &       0.24            \\      
                4.3     $\pm$   0.7     &       1.56    $\pm$   0.08    &       0.02    $\pm$   0.002   &       331.0   &       0.32            \\      
                1.1     $\pm$   0.2     &       1.18    $\pm$   0.10    &       0.001   $\pm$   0.003   &       128.6   &       0.20            \\      
                1.4     $\pm$   0.3     &       1.27    $\pm$   0.10    &       0.003   $\pm$   0.002   &       124.4   &       0.26            \\      \hline
                                                                                                                                                                                                                                                        
                                        &                               &       Obs.\,2                 &       dof=117 &                       \\      
                2.9     $\pm$   0.2     &       1.02    $\pm$   0.04    &       0.16    $\pm$   0.01    &       2003.6  &       0.25            \\      
                3.1     $\pm$   0.2     &       1.02    $\pm$   0.04    &       0.18    $\pm$   0.01    &       2286.8  &       0.26            \\      
                3.5     $\pm$   0.2     &       1.09    $\pm$   0.04    &       0.19    $\pm$   0.01    &       2130.3  &       0.25            \\      
                2.7     $\pm$   0.2     &       1.06    $\pm$   0.05    &       0.13    $\pm$   0.01    &       1553.2  &       0.24            \\      
                2.6     $\pm$   0.2     &       1.14    $\pm$   0.05    &       0.11    $\pm$   0.01    &       1397.4  &       0.25            \\      
                1.8     $\pm$   0.1     &       0.99    $\pm$   0.05    &       0.06    $\pm$   0.01    &       922.7   &       0.22            \\      
                1.9     $\pm$   0.2     &       1.17    $\pm$   0.06    &       0.05    $\pm$   0.01    &       669.0   &       0.21            \\      
                1.5     $\pm$   0.1     &       1.22    $\pm$   0.06    &       0.03    $\pm$   0.004   &       453.7   &       0.21            \\      
                2.0     $\pm$   0.2     &       1.27    $\pm$   0.05    &       0.03    $\pm$   0.004   &       471.6   &       0.21            \\      
                1.1     $\pm$   0.1     &       1.19    $\pm$   0.06    &       0.01    $\pm$   0.003   &       293.1   &       0.20            \\      
                1.2     $\pm$   0.1     &       1.28    $\pm$   0.07    &       0.01    $\pm$   0.003   &       233.6   &       0.14            \\      \hline
                                                                                                                                                                                                                                                        
                                        &                               &       Obs.\,3                 &       dof=117                         \\      
                2.3     $\pm$   0.2     &       0.80    $\pm$   0.04    &       0.08    $\pm$   0.01    &       2806.7  &       0.35            \\      
                2.2     $\pm$   0.1     &       0.71    $\pm$   0.04    &       0.06    $\pm$   0.01    &       3437.1  &       0.35            \\      
                2.8     $\pm$   0.2     &       0.85    $\pm$   0.04    &       0.10    $\pm$   0.01    &       3283.3  &       0.35            \\      
                2.3     $\pm$   0.2     &       0.87    $\pm$   0.04    &       0.06    $\pm$   0.01    &       2491.0  &       0.35            \\      
                2.0     $\pm$   0.2     &       0.90    $\pm$   0.04    &       0.05    $\pm$   0.01    &       1822.8  &       0.33            \\      
                1.8     $\pm$   0.2     &       0.96    $\pm$   0.05    &       0.04    $\pm$   0.01    &       1241.2  &       0.30            \\      
                1.1     $\pm$   0.1     &       0.82    $\pm$   0.05    &       0.01    $\pm$   0.01    &       922.1   &       0.30            \\      
                1.4     $\pm$   0.2     &       1.12    $\pm$   0.06    &       0.02    $\pm$   0.003   &       550.3   &       0.26            \\      
                2.1     $\pm$   0.2     &       1.21    $\pm$   0.05    &       0.01    $\pm$   0.003   &       545.0   &       0.28            \\      
                1.1     $\pm$   0.2     &       1.24    $\pm$   0.07    &       0.01    $\pm$   0.002   &       347.1   &       0.28            \\      
                1.1     $\pm$   0.2     &       1.21    $\pm$   0.07    &       0.004   $\pm$   0.002   &       268.6   &       0.24            \\      \hline
                                                                                                                                                                                                                                
                                        &                               &       Obs.\,4                 &       dof=107 &                       \\      
                3.4     $\pm$   0.2     &       0.77    $\pm$   0.03    &       -0.01   $\pm$   0.01    &       2815.9  &       0.58            \\      
                3.3     $\pm$   0.2     &       0.72    $\pm$   0.03    &       -0.03   $\pm$   0.01    &       3225.0  &       0.58            \\      
                3.2     $\pm$   0.2     &       0.73    $\pm$   0.03    &       -0.02   $\pm$   0.01    &       3015.6  &       0.55            \\      
                2.7     $\pm$   0.2     &       0.78    $\pm$   0.03    &       -0.01   $\pm$   0.01    &       7119.4  &       0.56            \\      
                2.4     $\pm$   0.2     &       0.80    $\pm$   0.03    &       -0.01   $\pm$   0.005   &       1804.5  &       0.56            \\      
                2.3     $\pm$   0.2     &       0.90    $\pm$   0.04    &       -0.001  $\pm$   0.004   &       1275.4  &       0.56            \\      
                1.6     $\pm$   0.2     &       0.88    $\pm$   0.04    &       -0.01   $\pm$   0.003   &       998.8   &       0.72            \\      
                1.3     $\pm$   0.2     &       1.00    $\pm$   0.05    &       -0.002  $\pm$   0.001   &       575.9   &       0.71            \\      
                2.0     $\pm$   0.4     &       1.23    $\pm$   0.07    &       0.004   $\pm$   0.001   &       301.7   &       0.42            \\      
                0.9     $\pm$   0.2     &       1.17    $\pm$   0.09    &       2.1E-4  $\pm$   0.001   &       245.9   &       0.51            \\      
                0.9     $\pm$   0.2     &       1.14    $\pm$   0.08    &       2.7E-5  $\pm$   0.001   &       198.6   &       0.54            \\      \hline
                                                                                                                                                                                                                                                        
                                        &                               &       Obs.\,5                 &       dof=117 &                       \\      
                3.2     $\pm$   0.1     &       0.79    $\pm$   0.02    &       0.004   $\pm$   0.01    &       1909.0  &       0.31            \\      
                3.6     $\pm$   0.1     &       0.80    $\pm$   0.02    &       -0.001  $\pm$   0.01    &       2173.3  &       0.32            \\      
                3.8     $\pm$   0.1     &       0.83    $\pm$   0.02    &       0.01    $\pm$   0.01    &       2203.7  &       0.32            \\      
                2.7     $\pm$   0.1     &       0.81    $\pm$   0.02    &       0.005   $\pm$   0.01    &       1553.7  &       0.30            \\      
                2.6     $\pm$   0.1     &       0.89    $\pm$   0.02    &       0.02    $\pm$   0.01    &       1291.3  &       0.29            \\      
                2.2     $\pm$   0.1     &       0.93    $\pm$   0.02    &       0.01    $\pm$   0.004   &       1070.9  &       0.33            \\      
                1.9     $\pm$   0.1     &       1.00    $\pm$   0.03    &       0.01    $\pm$   0.003   &       688.4   &       0.30            \\      
                1.5     $\pm$   0.1     &       1.09    $\pm$   0.03    &       0.01    $\pm$   0.002   &       441.8   &       0.29            \\      
                2.3     $\pm$   0.1     &       1.27    $\pm$   0.03    &       0.01    $\pm$   0.002   &       389.6   &       0.24            \\      
                1.4     $\pm$   0.1     &       1.32    $\pm$   0.04    &       0.004   $\pm$   0.001   &       279.8   &       0.27            \\      
                1.0     $\pm$   0.1     &       1.16    $\pm$   0.04    &       0.001   $\pm$   0.001   &       205.2   &       0.23            \\      \hline

\hline

\end{tabular}
\label{table:PLcparam}
\end{table}


\subsection{The spectral shape of the soft-band `constant component'}
\label{subsec:spectralfit}

As mentioned above, the non-zero, positive constants detected in the FFPs of Obs.\,2 and Obs.\,3, and perhaps Obs.\,1 as well, could be indicative of the presence of a separate spectral component which is less variable than the primary X-ray source on timescales of a few ks. In this case the best-fit  {\it c} model values can be used to study the spectral shape of the constant component. For that reason we divided them by the corresponding energy band width  to produce the 'spectrum` of this component, in ${\rm photon\,s^{-1}keV^{-1}}$. 

First, we fitted those spectra with an absorbed blackbody model ({\tt wabs$\ast$bbody}) by considering only the Galactic absorption. The best-fit $kT_{\rm bb}$ turned out to be similar when we fitted the three spectra individually, so we repeated the fit by keeping $kT_{\rm bb}$ tied in all of them. The resulting best-fit temperature was  $110\pm 1.8 \,{\rm eV}$, but the overall quality of the fit is poor  ($\chi^2/{\rm dof}=87.9/29$). Then we considered an absorbed power-law model ({\tt wabs$\ast$powerlaw}). The best-fit $\Gamma$ is  steep: $\sim 4$ for Obs.\,2 and Obs.\,3, and $\sim 4.5$ for Obs.\,1, but the overall quality of the fit is worse than before ($\chi^2/{\rm dof}= 137.7/27$). 

Then we fitted the spectra with the {\tt optxagnf} model \citep{Don12}.  This is a model for the spectral energy distribution of a disc around a rotating SMBH. 
We first fitted the spectra with {\tt wabs$\ast$optxagnf}, assuming Galactic absorption only, and setting {\it rcor} tied to the same value in all spectra ({\it rcor}, measured in ${\rm r_g}$, sets the radius below which the disc emission emerges as a low-temperature, large, optical-depth, Compton-upscattered flux, as opposed to a colour-temperature-corrected blackbody at larger radii). We ignored the emission of the hard X-ray corona, by keeping the model parameter {\it fpl} frozen to $10^{-6}$, since we did not take  the full band spectrum of the source into consideration. To constrain the fit as much as possible, we kept the black hole (BH) mass frozen to $10^7\,{\rm M_{\odot}}$ \citep[e.g.][]{Zho05,Emma14}. We also considered a maximally rotating BH by freezing the spin parameter, {\it astar}, to its maximum value of 0.998 in all spectra. The best-fit Eddington ratio $\left( \log(L/L_{\rm Edd})\right)$ turned out to be similar for Obs.\,1 and 3, and for that reason we repeated the fit by keeping it tied for these two observations. We also kept $kT_{\rm e}$ and $\tau$ (the electron temperature and optical depth for the soft Comptonization component) tied to the same values in all spectra. 

The quality of the fit improves significantly ($\chi^2/{\rm dof} = 52.6/28$), but this is still not a statistically accepted fit. We obtained a $3\sigma$ lower limit on the coronal radius of $rcor > 1.93\,{\rm r_g}$. The best-fit Eddington ratio is $\log(L/L_{\rm Edd})=-1.36^{+0.25}_{-0.20}$ for Obs.1 and 3, and $\log(L/L_{\rm Edd})=-1.10_{-0.22}^{+0.74}$ for Obs. 2.  Although the best-fit required different accretion ratios for the Obs.\,1 and 3 and Obs.\,2 spectra, the resulting best-fit values are consistent with each other within the errors. The best-fit electron temperature and optical depth were $kT_{\rm e} = 0.15_{-0.02}^{+0.04}{\,\rm keV}$ and $\tau = 22^{+3.0}_{-2.5}$, respectively. 

\subsection{The slope of the flux--flux plots}
\label{subsec:simulation}
\begin{figure}
\centering
\includegraphics[scale=0.45]{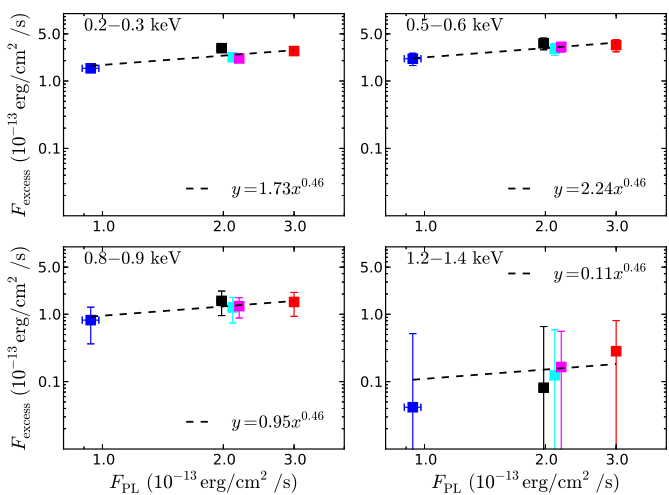}
\caption{$F_{\rm excess,obs}$ plotted as a function of $F_{\rm PL,obs}$, for Obs.1, 2, 3, 4, and 5 (black, red, cyan, blue, and magenta squares, respectively). The errors on the points were calculated using the {\tt efluxer} flux error and the error on the best-fit slope, and normalisation  of the PL model fit to the 1.7--3 keV data. The dashed lines indicate the best-fit PL model fits to the data. }
\label{fig:diff}
\end{figure}

The PLc best fit slope values of the FFPs at energies below $\sim 1$\,keV are less than unity in most observations. This result suggests the presence of intrinsic spectral variations in the continuum emission of the source. To investigate this issue further, we used the {\tt SAS} tool {\tt efluxer} and estimated the source flux (in ${\rm erg/cm^2/s}$) in all the energy bands of all observations. We also fitted the Obs.\,1-5 spectra in the 1.7--3\,keV band with an absorbed PL model. We computed the best-fit PL flux in the same bands and we calculated the difference between the observed and the PL model flux. This should be representative of the `intrinsic' soft-band flux alone, $F_{\rm excess,obs}$. Figure\,\ref{fig:diff} shows  $F_{\rm excess,obs}$ versus the 1.7--3 keV flux (which should be representative of the X-ray primary flux, $F_{\rm PL,obs}$, for the 0.2--0.3, 0.5--0.6, 0.8--0.9, and 1.2--1.4 bands. (These plots are representative of the plots in all  energy bands.) . The observed excess flux in the 1.2--1.4 keV band (and in general at energies above 1 keV) is almost an order of magnitude smaller than the one at energies below $\sim 1$ keV, and it is hard to constrain. In fact, it may even be zero (in agreement with Fig.\,\ref{fig:ratio},  which shows that the soft excess flux at energies above $\sim 1$ keV is minimal).

We fitted the data in all energy bands with a PL model, with the slope fixed to 0.46. This is the best-fit slope value we found when we fitted the data in panel (a) of Fig. 6 in \cite{Chi15}. These data show the relation between the 0.1--100 keV flux of the reflection component (which may be responsible for the soft excess) and the primary flux (in the same band) in IRAS 13224-3809. This PL model fits  the ``$F_{\rm excess,obs}$ vs $F_{\rm PL,obs}$'' data in all bands reasonably well.

We then produced synthetic energy spectra (using the {\tt XSPEC} command {\tt fakeit}), assuming a PL model and taking the Galactic absorption into account, in which the spectral slope of the continuum varies with the continuum flux according to $\Gamma =5.73N_{\rm PL}^{0.1}$. This is the best-fit model to the slope and normalisation of the PL fits to the 1.7--3 keV band data of each observation, and it it is in good agreement with the spectral variability law ($\Gamma \propto N_{\rm PL}^{0.08}$) found by \cite{Sob09} .We considered two cases for the model variability: (a) We first tried to reproduce the source behaviour during Obs.\,2, which shows the highest flux, which meant we considered a mean continuum flux equal to $F_{\rm PL, Obs2}$ and an $N_{\rm PL}$ variability of a factor of 30, similar to the observed one (the high-flux case hereafter). (b) We did the same, but tried to model the variability seen in Obs.\,4, which shows the lowest flux and an $N_{\rm PL}$ variation flux variation of $\sim 50$ (the low-flux case hereafter). The high-flux case implies $\Gamma$ variation between 1.95 and 2.95, while the spectral slope varies between 1.6 and 2.55 in the low-flux case.

\begin{figure}
\centering
\includegraphics[scale=0.45]{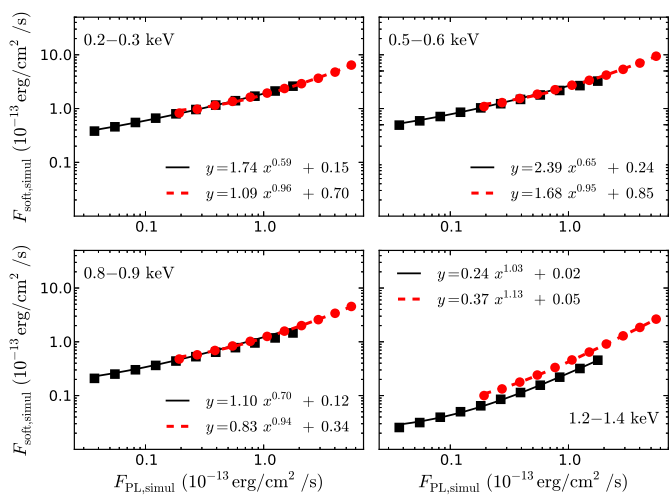}
\caption{Plots of $F_{\rm soft,simul}$ as a function of $F_{\rm PL,simul}$ in the high-flux and low-flux variability cases (red circles and black squares, respectively). The FPPs were fitted with a PLc model (dashed red  line and solid black  line for the high- and low-flux cases, respectively).}
\label{fig:simulation}
\end{figure}

We measured the flux of the simulated spectra in the soft-bands we consider in this work, as well as in the 1.7--3\,keV band ($F_{\rm PL,simul}$). We then added an `excess' flux to the soft-band fluxes, which we calculated using $F_{\rm PL,simul}$ and the best-fit PL models to the observed `$F_{\rm excess, obs}-F_{\rm PL,obs}$' plots in each band. Next, we plotted the total soft-band flux, $F_{\rm soft,simul}$, as a function of $F_{\rm PL,simul}$ for all the $N_{\rm PL}$ values. The resulting `synthetic' flux--flux plots are plotted in Fig.\,\ref{fig:simulation}. We fitted the simulated FFPs with a PLc model, exactly as we did with the observed FFPs.  The dashed and solid lines in Fig.\,\ref{fig:simulation} show these best-fit models (for the high- and low-flux cases, respectively). Interestingly, we found that the constant $c$ in the PLc model fits was not equal to zero. In other words, a flattening in the FFPs is expected, even when there is no constant soft component, but the soft-excess component is variable, albeit in a  fashion  correlated with the PL continuum.

The best-fit $\beta_{\rm PLc,mod}$ and $c_{\rm PLc,mod}$ model values are plotted in Fig.\,\ref{fig:PLc-Param}. To transform the best-fit $c_{\rm PLc,mod}$ values  (currently in ${\rm erg/cm^2/s}$ units) into count/s, we used the {\tt HEASOFT} web tool {\tt PIMMS}, assuming that the broad-band soft-excess component in IRAS 13224--3809 is  approximated  by a blackbody model, with $kT=0.1$\,keV. (Such a blackbody component does approximate the soft excess in this object in very broad terms, as we verified by fitting a PL plus {\tt bbody} model to the data shown in Fig.\,\ref{fig:ratio}.)

The high-flux case and low-flux case $\beta_{\rm PLc,mod}$ predictions bracket the observed values at energies below $\sim 0.9$ keV. At higher energies, both case models predict FFPs with slopes flatter than the observed ones. This is probably because the soft-excess contribution to the observed flux is minimal at high energies. For example, the empty diamonds and hexagons in the top panel of Fig.\,\ref{fig:PLc-Param} indicate the $\beta_{\rm PLc, mod}$ values when we only consider  the PL continuum with variable $\Gamma$, i.e. without adding $F_{\rm excess}$, for both cases. The agreement with the observed slopes is much better in this case. 

The model constants $c_{\rm PLc,mod}$ are smaller than the observed constants, both in the high- and low-flux cases. This indicates that, although a flattening at low fluxes is expected in the case of a primary variable in norm and shape and of a variable soft-band component, a stable component may still exist. The spectrum of this component can be fitted well with an optxagnf model ($\chi^2/{\rm dof}=12.7/10$) by fixing all the parameters to the best-fit values reported in Sec.\,\ref{subsec:spectralfit} for Obs.\,2 and only leaving   the $\log(L/L_{\rm Edd})$ as a free parameter. The new best-fit value for the Eddington ratio is $\sim 0.05$, which is $\sim 2$ times less than the previous value for this observation. 

\section{Discussion and conclusions}
\label{sec:Disc}

We have presented a timing analysis of the five {\it XMM-Newton} archival observations of the NLS1 Galaxy IRAS $13224-3809$. Our main aim was  to study its X-ray variability at energies below $\sim 1.5$\,keV. 

We produced FFPs using the data in 11 energy bands between 0.2 and 1.7\,keV and in the 1.7--3\,keV band data, which we assumed is representative of the X-ray primary `continuum'. We considered various bin sizes for the light curves, and we found that the bin size choice is important because it affects the shape of the resulting plots. This is probably the case in sources that show fast variations on short timescales, and where the intrinsic FFPs are non-linear. In these cases, the choice of the shortest possible bin size is preferred.

Our main results, using the 1-ks binned light curves, are summarised below:
\begin{itemize}

\item We fitted the FFPs with a PLc model. Positive constants are detected at energies below 1\,keV (where the soft excess is more pronounced in this source) in Obs.\,1, 2, and 3. All the best-fit values are consistent with zero in the case of Obs.\,5. We also detected negative constants in Obs.\,4, which are all
negative at all energies less than 1\,keV, although they are consistent with zero (at $3\sigma$).

\item The best-fit slopes are significantly flatter than 1 at energies below $\sim 1$ keV, except in Obs.\,2, where they are slightly steeper than unity. This indicates that the intrinsic flux--flux relation is not linear and  points to intrinsic spectral variability. 

\end{itemize}

Strictly speaking, the PLc model is not statistically accepted. Nevertheless, the model represents  the general trends in the flux--flux plots well. The residual plots do not show any large-scale, systematics residuals, which is indicative of the presence of an extra, broad-band model component. The high best-fit $\chi^2$ values are due to random data fluctuations around the best-fit models, which have an amplitude of $\sim 20-55\%$ (of the model value) in all observations. The observed range of variations in all observations is comparable to 5--10 in the soft energy bands. The PLc model, therefore, does take account of most of the observed variations, and there is just a scatter, of $\sim 0.2-0.5$, that is left in the residual's plots. This implies that  short-amplitude, fast variations in the  soft energy bands exist and do not depend on the hard band flux. However, the study of the causes of these variations is beyond the scope of the present work.

The flattening at low-flux rates that we detect in Obs.\,1, 2, and 3 at energies below $\sim 0.9-1$\,keV  could imply the presence of a separate component that is not variable on timescales of a few ks. However, the resulting spectrum of this  component cannot be fitted by either a blackbody, power-law, or by optxagnf \citep[which is a more realistic model for the accretion disc emission in AGN (][]{Don12}). 

For almost all observations, the best-fit PLc slopes are not equal to unity at all energies. This implies significant intrinsic spectral variations. We investigated the case of a primary continuum, which is variable in flux, and spectral slope as $\Gamma\propto N_{PL}^{0.1}$ \citep[in agreement with][]{Sob09} as well as a soft band excess flux, which is variable and which positively correlated with the continuum according to the relation $F_{\rm excess} \propto F_{\rm primary}^{0.46}$. We find that the slope of the resulting model FFPs depends on energy in a similar way to what we observed. The slope of the model FFPs at energies above $\sim 0.9\,{\rm keV}$ is similar to the slope of the observed plots, but only if the soft-band component does not contribute to these energies. In addition, the model FFPs flatten at low fluxes, resulting in the detection of a positive constant when fitted by a PLc model, although no constant soft-band component is present. 

The constant values that we obtained from the simulated FFPs are lower than the ones we observed. This may indicate the presence of a separate, constant component. The 0.2--1 keV band flux of this component should be $\sim 15$\% of the total observed flux during Obs.\,2. In the case of Obs\,2, however, this component is broadly consistent with the energy spectrum emitted by an accretion disc around a maximally rotating BH of mass $\sim 10^7\,{\rm M_{\odot} }$ with an accretion rate of $\sim 0.05$ of the Eddington limit. This is a rather low value for NLS1 galaxies, which are believed to accrete at much higher rates. At  radii $r < r_{\rm corona}=2 {\rm r_g}$ (at least), the energy is dissipated by Compton up-scattering of seed photons at the disc temperature at $r=r_{\rm corona}$, off electrons with a temperature of $\sim 0.15$\,keV and optical depth of $\sim 22$. However, the extrapolation of best-fit optxagnf model to the ultraviolet band  at $2310\,\AA$ gives a flux of $1.2\times 10^{-16}\,{\rm erg\,cm^{-2}\,s^{-1}\AA^{-1}}$ in the case of Obs.\,2. This is just $\sim 4$\% of the observed OM UVM2 flux, which is $3.07\times10^{-15}\,{\rm erg\,cm^{-2}\,s^{-1}\AA^{-1}}$, as reported by \cite{Chi15}. 

For this reason, we believe that the positive constants we detect in the FFPs are mainly due to the intrinsic spectral variations. Despite the agreement between the model and the observed FFPs' slope, it is difficult to explain the non-detection of positive constants in the Obs.\, 5 FFPs. Even if all the positive constants are caused by intrinsic, complicated spectral variations, their non-detection in Obs.\,5 implies that the source operates in a different way in this case (and yet the flux--flux slopes are similar to those observed in the other observations). It is even more difficult to explain the negative constants that we observe in the Obs.\,4 FFPs. One possibility is that the source is affected, at least occasionally, by extra, intrinsic absorption. This cannot be neutral; an extra neutral absorber can result in $\sim$ zero, but not negative, constants. In addition, the slopes will be significantly flatter than the observed ones in the low energy FFPs. On the other hand, an ionized absorber that is flux-dependent, in such a way that the absorption is stronger at low fluxes, may be able to explain the presence of negative constants in the FFPs. In practice, a warm absorber can vary in column density, covering fraction, and/or ionization parameter. The study of this kind of absorber is beyond the scope of the present work, since the modelling of its effects is highly complicated and not well constrained if all the parameters of the absorber are variable. 

In summary, our results support the hypothesis that most of the soft excess flux at energies below $\sim 0.9$ keV is due to X-ray reflection in IRAS 13224--3809. This  agrees with the results of \cite{Chi15} and with those of  \cite{Emma14} and \cite{Kar13}, who interpret the observed soft band time-lags in the same scenario. The soft excess is responding to the primary X-ray variations, although with a smaller amplitude (as expected for a smeared component). At the same time, the primary slope steepens with increasing flux \citep[which is again  agrees with the results of][]{Chi15}. We cannot exclude the presence of an extra, stable soft-band component, which can be well fitted by a disc plus a warm Comptonization medium emission. Its contribution to the observed flux, and at energies below 1 keV, should be less than $\sim 15$\%. 

The study of the FFPs can yield interesting results, not only when positive constants are detected (which may not correspond to a constant component), but also when none or even negative constants are detected. This is the case with two observations of IRAS 13224--3809, and they  may indicate the presence of an intrinsic warm and variable absorber in the source. 

\begin{acknowledgements}
This work is based on observations obtained with {\it XMM-Newton}, an {\it ESA} science mission with instruments and contributions directly funded by {\it ESA Member States} and {\it NASA}. We thank the anonymous referee for comments. ESK thanks G. Ghirlanda, A. Obuljen, and G. Oganesyan for discussions . :-)\end{acknowledgements}
 

\bibliographystyle{aa} 
\bibliography{kammoun-references} 

\end{document}